\documentclass[12pt]{iopart}

\usepackage{iopams}
\usepackage{graphicx,amssymb}
\usepackage{color}
\usepackage[utf8]{inputenc}
\newtheorem{lemma}{Lemma}
\newtheorem{proposition}{Proposition}
\newtheorem{theorem}{Theorem}

\newtheorem{definition}{Definition}

\newcommand{\be}{\begin{equation}}
\newcommand{\ee}{\end{equation}}
\def\tr{\mathop{\rm tr}\nolimits}

\def\dif{{\rm d}}
\def\ci{\mathop{\textrm{i}}\nolimits}

\def\Ric{{\rm Ric}}

\def\ci{{\rm i}}

\begin{document}

\title[Spatially-Homogeneous Cosmologies]
{Spatially-Homogeneous Cosmologies}

\author{Juan Antonio S\'aez$^1$, Salvador Mengual$^{2}$\footnote{Author to whom any correspondence should be addressed.} and Joan Josep Ferrando$^{2,3}$}

\address{$^1$\ Departament de Matem\`atiques per a l'Economia i l'Empresa,
Universitat de Val\`encia, E-46022 Val\`encia, Spain}

\address{$^2$\ Departament d'Astronomia i Astrof\'{\i}sica, Universitat
de Val\`encia, E-46100 Burjassot, Val\`encia, Spain}

\address{$^3$\ Observatori Astron\`omic, Universitat
de Val\`encia, E-46980 Paterna, Val\`encia, Spain}

\ead{juan.a.saez@uv.es; salvador.mengual@uv.es; joan.ferrando@uv.es}

\begin{abstract}
The necessary and sufficient conditions for a perfect fluid solution to define a spatially-homogeneous cosmology are achieved. These conditions are Intrinsic, Deductive, Explicit and ALgorithmic, and they offer an IDEAL labeling of these geometries. When a three-dimensional group acts on the three-dimensional space-like orbits, the Bianchi type of the model is also obtained.
\end{abstract}
%

%

\today

\section{Introduction}

The Bianchi cosmological models (spatially-homogeneous perfect fluid  solutions) gene\-ralize the Friedmann-Lema\^itre-Robertson-Walker solutions, and the Einstein equations also become an ordinary differential system for them. However, they are compati\-ble with anisotropies produced by rotation or global magnetic fields, and thus they can be more suitable to model the real universe, or to approximate it in some regions. 

The study of the Bianchi models began in the fifties \cite{taub, schucking} and it had a significant boom in the  late sixties \cite{E-W-Behr, Ellis-McCallum}. An exhaustive bibliography of that period can be found in the Ryan-Shepley's book \cite{RyanShepley}. More recent manuals present the basic concepts and references on the subject \cite{Ellis-Elst, Kramer, Krasinski-Plebanski, ellis-maar-mac}. The understanding of the Bianchi models took an important step forward with the introduction of their Hamiltonian formulation \cite{Uggla-1995} and the use of the Hubble-normalized variables \cite{Wainwright-1989}. These approaches have made it possible to study the asymptotic behavior of many cosmological models of different Bianchi types (see, for example, \cite{Hewitt-2003, Horwood-2003, Hervik-2008, Hervik-2010} and references therein).  The study of new solutions and the analysis of the geometric properties of the Bianchi models are currently ongoing (see, for example, the recent papers \cite{thorsrud, singh, parnovsky} and references therein).
 
An issue about the Bianchi cosmological models and the other spatially-homogeneous cosmologies that is still to be studied is their IDEAL characterization, that is, their labeling through conditions that are Intrinsic (depending only on the metric tensor), Deductive (not involving inductive or inferential methods or arguments), Explicit (the solutions are not expressed implicitly) and ALgorithmic (giving the solution as a flow chart with a finite number of steps). The aim of this paper is to obtain the aforementioned IDEAL characterization. This study requires considering different classes of perfect fluid solutions and, for each class, the necessary and sufficient conditions for a solution to define a spatially-homogeneous cosmology are given.

Recently \cite{SMF-ST} we have presented a first necessary step for implementing this approach: the IDEAL labeling of the spacetimes admitting a maximal group of isometries of dimensions three or four. These results are the starting point of the present paper and they enable us to study the spatially-homogeneous cosmologies admitting a {\it Riemann-frame} (a frame that can be built from the Riemann tensor and its covariant derivatives). 

Once we know that the spacetime is a spatially-homogeneous perfect fluid solution, a second step consists in determining the Bianchi type when a G$_3$ acting on a S$_3$ exists \cite{bianchi}. For this task we make use of the previous results obtained in  \cite{FS-G3}, where the different three-dimensional homogeneous Riemannian spaces have been intrinsically characterized. In our analysis, we write the classification that follows from the Sch\"{u}cking-Kundt-Behr approach \cite{E-W-Behr, Ellis-McCallum, Kramer, Krasinski-Plebanski, kundt} in terms of the Ricci tensor.

The IDEAL characterization of spacetimes provides an algorithmic way to test if a metric tensor, given in an arbitrary coordinate system, is a specific solution of Einstein equations. The conceptual interest and possible applications of this kind of studies has been widely pointed out elsewhere \cite{fs-SSST, fs-Szafron-Ideal}. The fundamental results by Cartan \cite{Cartan} and a wide understanding of the algebraic structure of the Ricci and Weyl tensors \cite{bcm, FMS-Weyl} play an essential role in carrying out the IDEAL labelling for any given metric. 

By applying these outcomes, we have performed an approach that has been useful in characterizing, among others, the Schwarzschild \cite{fs-S}, Reissner–Nordström \cite{fsD} and Kerr \cite{fs-Kerr} black holes, the Lemaître–Tolman \cite{fs-SSST} or Bertotti–Robinson \cite{fswarped} solutions, and the Stephani or the Szekeres–Szafron universes \cite{fs-SSST-Ricci, fs-Szafron-Ideal}. Other authors also have presented IDEAL approaches for higher dimensional spaces \cite{Khavkine, Khavkine-b}. These studies have been suitable in obtaining a fully algorithmic labelling of the initial data which correspond to a given solution \cite{garcia-parrado-vk, garcia-parrado, garcia-parrado-2016}. Recently, our results have been used to characterize the four dimensional Lorentzian spacetimes conformally related to the Kerr vacuum solution \cite{GP-conformal-Kerr}.

In all these studies and in the present paper we use some tensor concomitants of the Riemann tensor to perform an approach that is very suitable for setting up algorithms with a reasonably low number of steps. Note that the Cartan \cite{Cartan} approach, which was adapted to a Lorentzian four-dimensional manifold by Brans \cite{brans} and Karlhede \cite{karlhede} (see also \cite{karlhede-maccallum}), is based on the determination of Riemann scalar invariants. The number of scalars that must be computed is formidable \cite{tomoda} and general algorithms implementing the Cartan-Karlhede approach are difficult to build.

This paper is organized as follows. 
In section \ref{sec-dimension}, we introduce the connection tensor $H$ associated with an orthonormal Riemann-frame $\{ e_a \}$. It is a fundamental concept for developing the present study and it has showed its usefulness in characterizing the dimension of the isometry group in a four-dimensional spacetime \cite{SMF-ST} and in a three-dimensional Riemannian  \cite{FS-G3, FS-K3} or Lorentzian space \cite{FS-L3}. We also summarize some results in \cite{SMF-ST} that we use in the following sections. 

In order to get an IDEAL labeling of the spatially-homogeneous geometries we must obtain the connection tensor $H$ in terms of the Ricci and Weyl tensors. In section \ref{sec-computing-H}, we present these expressions for the solutions of Petrov-Bel types I, II, III and N, and we state the characterization theorem for a perfect fluid solution. Section \ref{sec-typeD} is devoted to analyze the type D spacetimes admitting a Riemann-frame, in which many cases must be considered.

The results in the above sections enable us to undertake in section \ref{sec-regular-SHC} the ideal labelling of the spatially-homogeneous cosmologies admitting a Riemann-frame. We analyse separately the cases when the maximal group is a G$_3$ and when the spacetime is homogeneous with a G$_4$ and admits a subgroup G$_3$ acting on three-dimensional space-like orbits. In both cases, the Bianchi type is also intrinsically characterized.

In section \ref{sec-singular-SHC}, we study the spatially-homogeneous cosmologies without a Riemann-frame. They are, either of Petrov-Bel type O, or type D and belong to the family with shear-free geodesic null principal directions. The case when the isometry group has three-dimensional orbits requires a more detailed analysis that is carried out in \ref{apen-typeD-nonconstant}.  

The IDEAL nature of our results enables us to present them in an algorithmic form in section \ref{sec-algoritmes} through several flow diagrams. In section \ref{sec-exemples} we illustrate the usefulness of our study by applying it to two examples: the Bianchi models of type I, and the homogeneous unimodular solutions of classes II and III by Ozsv\'ath \cite{Ozsvath_b, Farnsworth-Kerr}. In this second case we correct some mistakes that we have found in the literature.

Finally, in section \ref{sec-discussion} we comment about our results, the open problems that they suggest and their possible applications.


\section{Dimension of the isometry group when an invariant frame exists} 
\label{sec-dimension}

Let us consider a spacetime with metric $g$ of signature $\{-+++\}$ and metric volume element $\eta$. We shall denote with the same symbol a tensor and the metric  equivalent tensors that follow by raising and lowering indexes with $g$. If a Riemann-frame exists, we can always orthonormalize it and then work with such oriented orthonormal frame $\{ e_a \}$, $\eta = e_0 \wedge e_1 \wedge e_2 \wedge e_3$. Let $\{ \theta^{a} \}$ be its algebraic dual basis. The connection coefficients $\gamma_{ab}^c $ are defined as usual, $\nabla e_{a} = \gamma_{ab}^{c} \, \theta^{b} \otimes e_{c}$. We can collect the connection coefficients in the connection tensor $H$ \cite{SMF-ST}:
\begin{equation} \label{defH}
H= \gamma^{c}_{a b} \, \theta^{b} \otimes \theta^a \otimes e_{c} = - \frac12 \tilde{\eta}^{a b}\nabla e_{a} \bar{\wedge} e_b  \, ,
\end{equation}
where $\tilde{\eta}^{ab}$ is the signature symbol, $\tilde{\eta}^{ab}= diag(-1,1,1,1)$, and where, for a vector $v$ and a two-tensor $A$, $(A \bar{\wedge} v)_{\alpha \beta \gamma} = A_{\alpha \beta} v_\gamma \!- \! A_{\alpha \gamma} v_\beta$.

In what follows, we will use Latin indexes to count the vectors of the frame, and to indicate the components of a tensor in this non-holonomic frame, while we will use Greek indexes to indicate components in a coordinate frame.

The tensor $H$ provides the covariant derivative of the vectors $e_{a}$  by contracting the second index of $H$ with $e_{a}$, that is, $\left( \nabla e_{a} \right)_{\lambda \mu} =H_{\lambda \nu \mu} (e_{a})^\nu$.
If $\xi$ is a Killing field that leaves the frame $\{ e_a \}$ invariant, we have:
\begin{equation} \label{nabla-killing}
\nabla \xi = i(\xi) H \, ,
\end{equation}
where $i(\xi)t$ represents the interior product of a vector field $\xi$ with a $p$-tensor $t$.

The connection tensor $H$ has been used to determine the dimension of the isometry group when an invariant frame exists \cite{SMF-ST}. The result relies in the fact that all the Riemann invariants depend on the connection coefficients $\gamma^a_{bc}$ and their successive directional derivatives along the invariant frame, namely
$e_{a_1}(\gamma^a_{bc})$, $e_{a_2} e_{a_1}(\gamma^a_{bc})$, and so on. As well as the connection coefficients are collected in the connection tensor, their directional derivatives of order $q$ can be grouped in a tensor $C^{[q]}$, which is a differential concomitant of $H$ \cite{SMF-ST}. Thus, $C^{[1]} = e_d (\gamma^a_{b c}) \ \theta^d \otimes \theta^c \otimes \theta^b \otimes e_a = C^{[1]}(H)$, is given by:
\begin{equation} \label{c1}
 C^{[1]}_{{\alpha} \mu \nu \sigma} = \nabla_{\alpha} H_{\mu\nu \sigma} + {H_{\alpha \mu}}^\rho H_{\rho \nu \sigma}  +
{H_{{\alpha} \nu}}^\rho  H_{\mu \rho \sigma} + {H_{{\alpha} \sigma}}^\rho  H_{\mu \nu \rho} \,  ,
\end{equation}
and $C^{[2]} = e_f e_d (\gamma^a_{b c}) \ \theta^f \otimes  \theta^d \otimes \theta^c \otimes \theta^b \otimes e_a = C^{[2]}(H)$ takes the expression:
\begin{equation} \label{c2}
\hspace*{-2.2cm} C^{[2]}_{\alpha_2 \alpha_1 \mu \nu \sigma} \!=\!
\nabla_{\alpha_2} C^{[1]}_{\alpha_1 \mu \nu \sigma} \!+\! {H_{\alpha_2
\alpha_1}}^{\rho} C^{[1]}_{\rho \mu \nu \sigma}\! +\! {H_{\alpha_2
\mu}}^{\rho} C^{[1]}_{\alpha_1 \rho  \nu \sigma}\! +\! {H_{\alpha_2
\nu}}^{\rho} C^{[1]}_{\alpha_1  \mu \rho  \sigma}\! +\! {H_{\alpha_2
\sigma}}^{\rho} C^{[1]}_{\alpha_1 \mu  \nu \rho } \, .
\end{equation}

In \cite{SMF-ST} we use the concomitants $C^{[q]} \equiv C^{[q]}(H)$, $q = 1,2,3,4$, to perform an algorithm providing the dimension of the admitted maximal group of isometries when a Riemann-frame exists. 
In the present work we are interested in characterizing the spatially-homogeneous cosmologies. These solutions can appear in different ways. The first case corresponds to the spacetimes admitting a maximal G$_3$ on space-like three-dimensional orbits. The second case appears when a homogeneous spacetime with a G$_4$ also admits a (non-maximal) G$_3$ acting on space-like three-dimensional orbits. In these two situations we can apply the results in \cite{SMF-ST}. But we must also analyze the
case when a non trivial isotropy group exists, which will appear when no Riemann-frame exists.

Now, we summarize the two results in \cite{SMF-ST} that characterize the spacetimes admitting maximal isometry groups G$_4$ and a G$_3$. Hereafter, we shall use $\eta(C^{[p]}\!,C^{[q]})$ to indicate the contraction of two indexes of the volume element $\eta$ with the first index of the tensors $C^{[q]}$.
\begin{proposition} \label{propo-G4}
Let us consider a spacetime metric admitting a Riemann-frame with associated connection tensor $H$ and $C^{[1]}$ given in {\em (\ref{c1})}. Then, it admits a maximal group {\em G}$_4$ of isometries if, and only if,
\begin{equation} \label{G4}
\hspace{-21.0mm} {\rm G}_4: \qquad \qquad  \qquad C^{[1]} \equiv C^{[1]}(H) =0.
\end{equation}
\end{proposition}
\begin{proposition} \label{propo-G3}
Let us consider a spacetime metric admitting a Riemann-frame with associated connection tensor $H$ and $C^{[1]}$ and $C^{[2]}$ given in {\em (\ref{c1})} and {\em (\ref{c2})}. Then, it
admits a maximal group {\em G}$_3$ of isometries if, and only if,
\be \label{G3}
\hspace{-21.0mm} {\rm G}_3: \qquad \qquad  C^{[1]}\neq 0, \quad \eta\,(C^{[1]}\!,C^{[1]}) =0, \quad \eta\,(C^{[1]}\!,C^{[2]} )=0 \, .
\ee
\end{proposition}

Once we know the dimension of the isometry group, we can determine the orbits and their causal character (space-like, time-like or null). The orbits are the $r$-surface orthogonal to the gradient of the $4-r$ independent invariant scalars, and these gradients appear in the first component of the concomitants $C^{[q]}$ involved in the characterization of the G$_r$ \cite{SMF-ST}. In fact, every Killing vector $\xi$ fulfills $i(\xi) C^{[q]} = 0$. 

Here we are interested in space-like three-dimensional orbits. Now, conditions in proposition \ref{propo-G3} hold and $C^{[1]}$ generates only one independent invariant scalar. Then, for any vector $v$ and any 2-form $V$, $m_\lambda = C^{[1]}_{\lambda \rho \mu \nu}v^\rho V^{\mu \nu}$ defines a direction which is collinear with the gradient of this scalar. Consequently, we obtain:
\begin{proposition} \label{propo-G3causal}
If a spacetime fulfills conditions of proposition {\em \ref{propo-G3}}, then it admits a transitive  group of isometries G$_3$, and the orbits are the hypersurfaces orthogonal to the vector $m_\lambda = C^{[1]}_{\lambda \rho \mu \nu}v^\rho V^{\mu \nu}$, where $v$ is a vector and $V$ a two-form such that $m \not=0$. The orbits are space-like when $m^2 <0$.
\end{proposition}
%


\section{Computing the connection tensor $H$}
\label{sec-computing-H}

If a Riemann-frame $\{ e_{a} \}$ is explicitly known, expression (\ref{defH}) can be used to compute $H$ and then the results of the propositions above apply. But the explicit determination of the frame is not always necessary, and $H$ can be obtained in terms of an adequate invariant tensor. The existence of such a frame strongly relies on the geometric properties of the curvature tensor and, more specifically, on the geometric elements associated with the Weyl and the Ricci tensors. 

In spacetimes of Petrov-Bel types I, II or III, the Weyl tensor algebraically defines a principal frame. In \cite{FMS-Weyl} we have presented an algorithmic way to determine this Weyl frame. Moreover, in  \cite{SMF-ST} we have obtained the connection tensor $H$ for each of those Petrov-Bel types from some specific concomitants of the Weyl tensor. Now, we can use the cosmological observer $u$ to obtain $H$ for a Petrov-Bel type N spacetime, and to get an alternative method for types II and III.

On the other hand, if we want to label the spatially-homogeneous cosmologies we must use an invariant characterization of the perfect fluid solutions. Concerning this, we have the following result \cite{fs-SSST-Ricci}:
\begin{theorem} \label{theo-fluper-energy}
Consider the following concomitants of the Ricci tensor $R$:
\begin{equation}
\hspace{-10mm} r \equiv \tr R  , \qquad   S \equiv R - \frac14 r \, g  , \qquad q \equiv - 2 \frac{\tr S^3}{\tr S^2} , \qquad Q \equiv S - \frac14 q \, g . \label{fluper-definitions}
\end{equation}
A spacetime is a perfect fluid solution if, and only if, the Ricci tensor $R$ satisfies:
\begin{equation} \label{fluper-conditions-A2}
Q^2 +  q Q = 0  , \qquad  q \, Q(v,v) > 0   ,
\end{equation}
where $v$ is any time-like vector.
Moreover, the energy density $\rho$, the pressure $p$ and the unit velocity $u$ of the fluid are given by:
\begin{equation} \label{fluper-hydro}
\hspace{-10mm} \rho = \frac14 (3 q + r)   , \qquad p = \frac14 (q - r)  , \qquad  u =  \frac{P(v)}{\sqrt{P(v,v)}} , \qquad P \equiv \frac{1}{q} Q  .
\end{equation}
\end{theorem}

We consider in this paper some cases where the orthonormal Riemann-frame $\{u, e_A\}$, $A=1,2,3$, is defined by a unit time-like vector $u$ and the three eigenvectors of a symmetric 2-tensor $E$, which is orthogonal to $u$ and admits three different eigenvalues. In \cite{SMF-ST} we have shown: 
\begin{lemma} \label{lemma-uE}
Let $u$ be a unit time-like direction and $E$ a symmetric traceless {\em 2}-tensor with $\delta \equiv (\tr E^2)^3-6(\tr E^3)^2 \neq 0$ and $E(u)=0$. The connection tensor $H$ associated with the frame defined by $u$ and the eigenframe of $E$ can be obtained as:
\be
\hspace{-20mm} H= \nabla u \bar{\wedge} u - J  , \qquad J_{\alpha \gamma \lambda} \equiv \nabla_\alpha E^{\mu}_{\ \rho} \, E^{\rho \nu} \,  \epsilon_{\mu\nu \pi} K^{\pi \sigma} \epsilon_{\sigma \gamma \lambda} , \quad \epsilon_{\sigma \gamma \lambda} \equiv  \eta_{\sigma \gamma \lambda \delta} u^{\delta}  , 
\ee
\be
\hspace{-20mm}  K \equiv \frac{1}{\delta} [ 3 \alpha E^2 + 6 \beta E + \frac{1}{2} \alpha^2 (g\!+\! u \otimes u)] \, , \quad  \alpha \! \equiv \! \tr E^2, \ \ \beta\! \equiv \! \tr E^3 , \ \ \delta\! \equiv \! \alpha^3 - 6 \beta^2   .
\ee
\end{lemma}
The unit vector $u$ could be defined by the gradient of a non-constant Riemann invariant scalar or it could be the cosmological observer. In this last case, it can be obtained form the Ricci tensor as (\ref{fluper-hydro}), an expression where an arbitrary vector $x$ appears. We can avoid the use of this $x$ in the lemma above because all the tensorial functions involved are quadratic in the unit vector $u$, and we can obtain them by using the projector $P=u \otimes u$ given in (\ref{fluper-hydro}). More precisely, we have:
\begin{lemma} \label{lemma-PE}
Let $P$ be the projector on a unit time-like direction $u$, $P= u \otimes u$, and $E$ a symmetric traceless {\em 2}-tensor with $\delta \equiv (\tr E^2)^3-6(\tr E^3)^2 \neq 0$ and $E\cdot P=0$. The connection tensor $H$ associated with the frame defined by $u$ and the eigenframe of $E$ can be obtained as:
\be
\hspace{-20mm} H= I - J  , \quad I_{\alpha \gamma \lambda} \equiv - \nabla_{\alpha}
{P^{\rho}}_{[\gamma} P_{\lambda] \rho} , \quad J_{\alpha \gamma \lambda} \equiv \nabla_\alpha E^{\mu}_{\ \rho} \, E^{\rho \nu} \,  \eta_{\mu \nu \pi \delta} K^{\pi \sigma} \eta_{\sigma \gamma \lambda \beta} P^{\delta \beta}  ,  \label{H-lemma2a}
\ee
\be
\hspace{-20mm}  K \equiv \frac{1}{\delta} [ 3 \alpha E^2 + 6 \beta E + \frac{1}{2} \alpha^2 (g\!+\! P)] \, , \quad  \alpha \! \equiv \! \tr E^2, \ \ \beta\! \equiv \! \tr E^3 , \ \ \delta\! \equiv \! \alpha^3 - 6 \beta^2   . \label{H-lemma2b}
\ee
\end{lemma}
Note that the above expressions (\ref{H-lemma2a}-\ref{H-lemma2b}) become expressions (10-11) if we take $P = u \otimes u$.

In this section we use the notation adopted in \cite{FMS-Weyl} when getting certain conco\-mitants of the Weyl tensor. The self-dual Weyl tensor is ${\cal W}= \frac{1}{2} (W - \ci *W)$, and ${\cal G} = \frac{1}{2} (G- \ci \eta)$ denotes the induced metric on the space of the self-dual bivectors, $G_{\alpha \beta \lambda \mu} = g_{\alpha \lambda} g_{\beta \mu} - g_{\alpha \mu} g_{\beta \lambda}$. We denote ${\cal W}^2$ the double bivector $({\cal W}^2)_{\alpha \beta \mu \nu} = \frac{1}{2} {\cal W}_{\alpha \beta \lambda \rho} {{\cal W}^{\lambda \rho}}_{\mu \nu}$, and $\Tr {\cal W} = \frac{1}{2}  {{\cal W}^{\mu \nu}}_{\mu \nu}$.


\subsection{Type I spacetimes}
\label{subsec-typeI}

In a type I spacetime, the self-dual Weyl tensor ${\cal W}$ is algebraically general and a Weyl principal frame of vectors $\{e_a\}$ exists \cite{FMS-Weyl}.  Then, we have the following result that has been shown in \cite{SMF-ST}:
\begin{proposition} 
\label{propo-typeI}
Let ${\cal W}$ be a type I Petrov-Bel Weyl tensor. The connection tensor $H$ associated with the Weyl principal frame can be obtained as
\be \label{H-Prop3}
H =\frac{1}{\sqrt{2}}({\cal H} + \bar{\cal H}) \, ,  \qquad {\cal H}_{\alpha   \mu \nu} \equiv \frac{1}{\sqrt{2}} {\cal X}_{\alpha \lambda \rho} {{\cal Y}^{\lambda \rho}}_{\mu \nu} \, ,
\ee
\be
{\cal X}_{\lambda \alpha \rho} \equiv \frac{1}{2} \nabla_{\lambda} {\cal W}_{\alpha \beta \mu \nu}  {{\cal W}^{\mu \nu \beta}}_{\rho}\, , \qquad {\cal Y} \equiv \frac{1}{\Delta} (3 a {\cal W}^2 + 6 b {\cal W} +
\frac{1}{2} a^2 {\cal G} ) \, ,
\ee
where $a\equiv \Tr {\cal W}^2$, $b\equiv \Tr {\cal W}^3$ and $\Delta \equiv a^3 - 6 b^2 \not= 0$.
\end{proposition}
%


\subsection{Spacetimes of types N, III or II}
\label{subsec-N-III-II}

Let ${\cal N}\not=0$ be a self-dual Weylian tensor of Petrov-Bel type N, ${\cal N}^2=0$. Then, ${\cal N} ={\cal L} \otimes {\cal L}$, where ${\cal L}$ is a null bivector \cite{FMS-Weyl}. If a unit time-like vector $u$ is given, we have that ${\cal L}= \ell \wedge (e_2 + \ci e_3 )$, $\ell$ being the null fundamental direction of ${\cal L}$ and $\{e_2, e_3\}$ an orthonormal basis orthogonal to
both $\ell$ and $u$. Then, if $P = u \otimes u$, we have that
\be
{\cal N}[P] = (\ell ,u)^2 ( e_2 \otimes e_2 - e_3 \otimes e_3 + \ci \, e_2 \widetilde{\otimes} e_3) ,
\ee
where, hereinafter, for a symmetric tensor $P$ and a double 2-form ${\cal N}$ we shall denote
${\cal N}[P]_{\alpha \mu} = {\cal N}_{\alpha \beta \mu \nu} \, P^{\beta \nu}$. 

This way, $E\equiv ({\cal N}+ \bar{{\cal N}})[P]$ is the electric part of ${\cal N}$ with respect to $u$ and it defines a traceless
tensor $E= 2(\ell,u)^2 (e_2 \otimes e_2 - e_3 \otimes e_3)$, which is orthogonal to $u$ and algebraically general. Then, it fulfils the conditions imposed in lemma \ref{lemma-PE}.

On the other hand, we can associate a self-dual Weylian tensor ${\cal N}$ with a Weyl tensor of type III or II \cite{FMS-Weyl, Debever}. Then, we have the following:
\begin{proposition}  \label{propo-N-III-II}
Let ${\cal W}$ be the self-dual Weyl tensor of a perfect fluid solutions of Petrov-Bel type N, III or II, and $R$ its Ricci tensor.  The connection tensor $H$ associated with a Riemann-frame can be obtained as in lemma {\em \ref{lemma-PE}}, with $P$ given in {\em (\ref{fluper-hydro})} and $E=({\cal N} + \bar{\cal N})[P]$, where:

i) ${\cal N}={\cal W}$ if the Weyl tensor  is of type N.

ii) ${\cal N}= {\cal W}^2$ if the Weyl tensor is of type III.

iii) ${\cal N} = {\cal W}^2 + \psi {\cal W} - 2 \psi^2 {\cal G}$, with
$\psi=- \frac{\tr {\cal W}^3}{\tr {\cal W}^2} $, if the Weyl tensor
is of type II.
\end{proposition}
%


\subsection{Spacetimes of types O and D}
\label{subsec-O-D}

The study of the spatially-homogeneous cosmologies of Petrov-Bel type D leads to two situations. When a Riemann frame exists (regular type D solutions), getting the associated connection tensor implies to analyze a large amount of cases, which will be considered in the following section. 
The singular type D cosmologies (no Riemann frame exists) will be analyzed in subsections \ref{subsec-TypeD-constant} and \ref{subsec-TypeD-nonconstant}.

On the other hand, no conformally flat spatially-homogeneous cosmology admits a Riemann-frame, and this case will be analyzed in subsection \ref{subsec-TypeO}.


\section{Computing $H$ in regular type D solutions}
\label{sec-typeD}

The self-dual Weyl tensor ${\cal W}$ of a type D spacetime admits the canonical form \cite{FMS-Weyl}:
\begin{equation} \label{Weyl-D}
{\cal W} = \psi \,( 3\, {\cal U} \otimes {\cal U} + {\cal G}),
\qquad \psi \equiv -\frac{\tr {\cal W}^3}{\tr{\cal W}^2} = \psi_R + \ci \psi_I \, ,
\end{equation}
where $\psi$ is the double Weyl eigenvalue and  ${\cal U}= \frac{1}{\sqrt{2}} (U - \ci *U)$, $U= \ell_+ \wedge \ell_-$ being the volume element of the time-like principal plane and $\ell_{\pm}$ the two double Debever directions. 

Let $u$ be the cosmological observer and $P$ the projector defined in (\ref{fluper-hydro}). Then, the electric part of the Weyl tensor is $E_1 \equiv W[P]$. A straightforward calculation shows that $E_1$ is algebraically general if, and only if, $u$ does not define a principal direction, that is, it is not on the principal plane. This fact can be characterized by the non-nullity of the scalar:
\be \label{delta_1}
\delta_1 \equiv (\tr E_1^2)^3-6(\tr E_1^3)^2 \, , \qquad  E_1 \equiv W[P] \, .
\ee
Then, if $\delta_1 \not= 0$, $E_1$ is a traceless algebraically general tensor and lemma \ref{lemma-PE} applies.

\begin{proposition} \label{propo-caso1} 
Let $W$ be the Weyl tensor of a perfect fluid solution of Petrov-Bel type D and $R$, its Ricci tensor. Let $P$, $E_1$ and $\delta_1$ be the Riemann concomitants given in {\em (\ref{fluper-hydro})} and {\em (\ref{delta_1})}. If $\delta_1 \not=0$, the connection tensor $H$ associated with a Riemann-frame can be obtained as in lemma {\em \ref{lemma-PE}}, with $E= E_1$.
\end{proposition}

From here on, we have to consider the case of $u$ being in the principal plane of the Weyl tensor, that is, $\delta_1=0$. In what follows, we denote $e_1$ the unit space-like principal direction orthogonal to $u$, $U=u\wedge e_1$. Then, the projector on the direction $e_1$ can be obtained as:
\begin{equation} \label{P_1}
e_1 \otimes e_1 =P_1 \, , \qquad    P_1 \equiv \frac{2}{3} \Big[ \frac{1}{\psi} {\cal W} - {\cal G} \Big] [P] \, .
\end{equation}
At this point, we distinguish two different situations characterized by the existence or not of a non-constant algebraic scalar invariant. 


\subsection{The algebraic scalar invariants are constants}
\label{subsec-constant}

In a type D perfect fluid solution, we have two Ricci algebraic scalar invariants, namely $r$ and $q$ defined in (\ref{fluper-definitions}), and the complex Weyl eigenvalue $\psi$ given in (\ref{Weyl-D}). In this subsection we study the case where these invariants are constant:
\be
\dif r = \dif q = \dif \psi = 0 \, .  \label{drqpsi}
\ee
Then, the contracted Bianchi identities imply that $u$ is geodesic and expansion free. This means that $\nabla u = \sigma+ \Omega$, where $\sigma$ and $\Omega$ are, respectively, the shear and the rotation tensors of the observer $u$. We have $\nabla P \cdot P = - (\sigma+ \Omega) \otimes u$, and consequently, we can obtain the shear tensor and the associated rotation vector $\omega$ in terms of the projector $P$. More precisely, we obtain:
\begin{eqnarray} \label{rot} 
\omega = *(u \wedge \dif u) = 2*(u \wedge \Omega)  \, , \qquad  \omega^\alpha \equiv - \eta^{\alpha \beta \lambda \mu} \nabla_{\beta} P_{\lambda \nu} P^{\nu}_{\ \mu} \, ,  \\
\sigma \equiv [P(v,v)]^{-1/2} \Sigma(v)  \, , \qquad  \Sigma_{\alpha \beta \lambda} \equiv  -\frac{1}{2} [ \nabla_{\alpha} P_{\beta \mu} + \nabla_{\beta} P_{\alpha \mu} ] \, P^{\mu}_{\, \lambda} \, , \label{Sigma}
\end{eqnarray}
where $v$ is an arbitrary time-like vector. If the rotation vector $\omega$ is not collinear to $e_1$, $\omega \wedge e_1 \neq 0$, it defines a direction orthogonal to $u$ that can be used to determine a traceless algebraically general tensor:
\begin{equation} \label{e2}
E_2 \equiv \omega^2 \, P_1 - \omega \otimes \omega \, ,
\end{equation}
which fulfills the conditions in lemma \ref{lemma-PE}. 

Otherwise, if $\omega \wedge e_1 = 0$ (that is, $ E_2 = 0$), we can use the shear tensor $\sigma$ to obtain a Riemann-frame. If $\sigma$ is algebraically general, the scalar
\be \label{deltasigma}
\delta_{\sigma} \equiv (\tr \sigma^2)^3 - 6 (\tr \sigma^3)^2 ,
\ee
does not vanish, and then we can apply lemma \ref{lemma-PE} with $E= \sigma$. 
On the contrary, if $\sigma \neq 0$ and $\delta_\sigma =0$, then it is algebraically special and takes the expression $\sigma= s (e_3  \otimes e_3- \gamma/3)$, where $e_3$ is the unit eigenvector associated with the simple eigenvalue $2s/3 \not=0$. Moreover, we obtain:
\begin{equation} \label{e3}
e_1 \otimes e_1 - e_3 \otimes e_3 = E_3 , \qquad E_3 \equiv P_1 - \frac{1}{s} \sigma - \frac{1}{3}   \gamma \, .
\end{equation}
Note that if $e_3 \neq e_1$, then $E_3 \not=0$ is a traceless algebraically general tensor and lemma \ref{lemma-PE} applies. The results in this subsection can be summarized in the following:
\begin{proposition} \label{propo-constant} 
Let $W$ be the Weyl tensor of a perfect fluid solutions of Petrov-Bel type D and $R$, its Ricci tensor. Let us assume that $\delta_1= 0$ and that equation {\em (\ref{drqpsi})} holds, where $r$, $q$, $\psi$ and $\delta_1$ are given in {\em (\ref{fluper-definitions}), (\ref{Weyl-D})} and {\em (\ref{delta_1})}. Let us consider the Riemann concomitants $P$, $P_1$, $\omega$, $\sigma$, $E_2$ and $\delta_\sigma$ given in {\em (\ref{fluper-hydro})} {\em (\ref{P_1})}, {\em (\ref{rot})} {\em (\ref{Sigma})}, {\em (\ref{e2})} and {\em (\ref{deltasigma})}. Then, the connection tensor $H$ associated with a Riemann-frame can be obtained as in lemma {\em \ref{lemma-PE}} with:
\begin{itemize}
\item[(i)] 
$E= E_2$ if $E_2 \not=0$.
\item[(ii)] 
$E= \sigma$ if $E_2=0$ and $\delta_\sigma \not=0$.
\item[(iii)] 
$E= E_3$, if $E_2 =0$, $\sigma \not=0$, $\delta_\sigma=0$, and $E_3$ given in {\em (\ref{e3})} does not vanish.
\end{itemize}
\end{proposition}

After this result, we only need to analyze when $\sigma =0$, or $\sigma \not=0$ and $E_3=0$. Under these constraints, and taking into account the Bianchi identities, we obtain:
\begin{equation} \label{nablau-nablae}
\hspace{-22mm} \nabla u = s (e_1 \otimes e_1 \!-\! \frac13 \gamma) + w *U, \quad \ \nabla e_1 =  \frac{2}{3} s\, e_1 \otimes u   , \quad \ s \equiv 3 \frac{ \tr \sigma^3}{\tr \sigma^2}, \quad \  w^2 = \omega^2    .
\end{equation}
From these expressions, we can show that the principal null directions, $\ell_{\pm}= u \pm e_1$, are null shear-free geodesics, and then these metrics belong to the class considered by Stewart and Ellis \cite{Stewart-Ellis} (see also \cite{Kramer, Wainwright}). The spatially-homogeneous cosmologies in this family will be considered in subsection \ref{subsec-TypeD-constant}.


\subsection{A non-constant algebraic invariant scalar exists}
\label{subsec-nonconstant}

Now we consider that at least one of the algebraic invariants is a non-constant scalar whose gradient defines a time-like direction:
\begin{equation} \label{mu}
\exists \  \mu \in \{r, q, \psi_R, \psi_I\}, \qquad  (\dif \mu)^2 < 0 .
\end{equation}
If this condition holds, there could be a maximal isometry group G$_3$, with $n$, the normal to the orbits, given by:
\begin{equation}\label{ene}
n = [-(\dif \mu)^2]^{-1/2} \dif \mu .
\end{equation}

The electric part of the Weyl tensor with respect to $n$ is $E_4 \equiv W[n \otimes n]$. Moreover, $E_4$ is algebraically general if, and only if, $n$ does not define a principal direction, that is, it is not on the principal plane. This fact can be characterized by the non-nullity of the scalar:
\be \label{delta_4}
\delta_4 \equiv (\tr E_4^2)^3-6(\tr E_4^3)^2 \, , \qquad  E_4 \equiv W[n \otimes n] \, .
\ee
Then, if $\delta_4 \not=0$,  $E_4$ fulfills the conditions imposed in lemma \ref{lemma-uE}. 

From here on, we have to consider the case of $n$ being in the principal plane of the Weyl tensor, that is, $\delta_4=0$. In what follows we denote $\tilde{e}_1$ the unit space-like principal direction orthogonal to $n$, $U=n\wedge \tilde{e}_1$. Then, the projector on the direction $\tilde{e}_1$ can be obtained as:
\begin{equation} \label{tildeP_1}
\tilde{e}_1 \otimes \tilde{e}_1 =\tilde{P}_1 \, , \qquad    \tilde{P}_1 \equiv \frac{2}{3} \Big[ \frac{1}{\psi} {\cal W} - {\cal G} \Big] [n \otimes n] \, .
\end{equation}

The gradient of the invariant scalar $(\dif \mu)^2$ must be orthogonal to the orbits, that is, 
\begin{equation} \label{dmu2}
\dif[ (\dif \mu)^2 ] \wedge n = 0 .
\end{equation}
This necessary condition is equivalent to $n$ defining a geodesic vorticity-free congruence, and then the shear $\tilde{\sigma}$ and the expansion $\tilde{\theta}$ are the only non-zero kinematic coefficients of $n$. Moreover, they can be obtained as:
\begin{equation} \label{sigmatilde}
\tilde{\sigma} \equiv \nabla n - \frac{1}{3} \tilde{\theta}\, \tilde{\gamma},  \qquad \tilde{\theta} \equiv  \nabla \cdot n , \qquad 
 \tilde{\gamma} \equiv  g + n \otimes n .
\end{equation}

The shear $\tilde{\sigma}$ fulfills the conditions in lemma \ref{lemma-uE} if it is algebraically general, that is, when the scalar 
\begin{equation} \label{deltasigmatilde}
\tilde{\delta}_\sigma \equiv ( \tr \tilde{\sigma}^2)^3 - 6(\tr \tilde{\sigma}^3)^2 , 
\end{equation}
does not vanish. Otherwise, if $\tilde{\delta}_\sigma =0$, then $\tilde{\sigma}$ is algebraically special and takes the expression $\tilde{\sigma}= \tilde{s} (\tilde{e}_3  \otimes \tilde{e}_3- \tilde{\gamma}/3)$, where $\tilde{e}_3$ is the unit eigenvector associated with the simple eigenvalue $2\tilde{s}/3$. Moreover, we obtain:
\begin{equation} \label{e5}
\tilde{e}_1 \otimes \tilde{e}_1 - \tilde{e}_3 \otimes \tilde{e}_3 = E_5 , \qquad E_5 \equiv \tilde{P}_1 -\frac{1}{\tilde{s}} \tilde{\sigma}- \frac{1}{3}   \tilde{\gamma} , \quad \tilde{s} \equiv 3\frac{\tr \tilde{\sigma}^3}{\tr \tilde{\sigma}^2}  \, .
\end{equation}
Note that if $\tilde{e}_3 \neq \tilde{e}_1$, then $E_5 \not=0$ is a traceless algebraically general tensor and lemma \ref{lemma-uE} applies. The results in this subsection can be summarized in the following:
\begin{proposition} \label{propo-nonconstant} 
Let $W$ be the Weyl tensor of a perfect fluid solutions of Petrov-Bel type D and $R$, its Ricci tensor. Let us assume that $\delta_1= 0$ and that condition {\em (\ref{mu})} holds, where $r$, $q$, $\psi$ and $\delta_1$ are given in {\em (\ref{fluper-definitions}), (\ref{Weyl-D})} and {\em (\ref{delta_1})}. Let us consider the Riemann concomitants $n$, $E_4$ and $\delta_4$ given in {\em (\ref{ene})} and {\em (\ref{delta_4})}. Then:
\begin{itemize}
\item[(i)] 
If $\delta_4 \not=0$, the connection tensor $H$ associated with a Riemann-frame can be obtained as in lemma {\em \ref{lemma-uE}}, with $u=n$ and $E= E_4$.
\item[(ii)] 
If $\delta_4=0$,  and {\em (\ref{dmu2})} holds, let $\tilde{\sigma}$ and $\delta_\sigma$ be given in {\em (\ref{sigmatilde}) and (\ref{deltasigmatilde})}. The connection tensor $H$ associated with a Riemann-frame can be obtained as in lemma {\em \ref{lemma-uE}}, with $u=n$ and (ii-a) $E= \tilde{\sigma}$ if $\tilde{\delta}_\sigma \not=0$, (ii-b)  $E= E_5$ if $\tilde{\delta}_\sigma =0$ and $E_5$ given in {\em (\ref{e5})} does not vanish.
\end{itemize}
\end{proposition}

In the time-like principal plane, we have two invariant orthonormal frames, $\{u, e_1\}$ and $\{n, \tilde{e}_1\}$, connected by a boost defined by a hyperbolic angle $\phi$, namely, 
\begin{equation} \label{phi}
u = \cosh \phi \, n + \sinh \phi \, \tilde{e}_1 , \qquad \cosh \phi = \Gamma \equiv \sqrt{P(n,n)} \, .
\end{equation}
A necessary condition for the existence of a G$_3$ is that all the scalar invariants have a collinear gradient:
\begin{equation} \label{lambda}
\dif \lambda \wedge n = 0, \qquad  \forall \lambda \in \{ r, q, \Gamma, \psi,  \tilde{\theta}, \tilde{s} \} \, .
\end{equation}
When none of the conditions in proposition \ref{propo-nonconstant} hold, constraint (\ref{lambda}) and Bianchi identities lead to:
\begin{equation} \label{nabla-e1}
\hspace{-22mm} \nabla n = \tilde{s} \tilde{e}_1 \otimes \tilde{e}_1 + \frac13 (\tilde{\theta} \! - \! \tilde{s})
\tilde{\gamma}, \quad \nabla \tilde{e}_1 =(\tilde{s} \!+\!
\tilde{\theta}/3) \tilde{e}_1 \otimes n + s_1
(\tilde{\gamma}\! -\! \tilde{e}_1 \otimes \tilde{e}_1)  + w_1 *\!U , 
\end{equation}
where the scalars $s_1$ and $w_1$ fulfill $\dif s_1 \wedge n = \dif w_1 \wedge n = 0$. Moreover, a straightforward calculation shows that no new directions can be obtained from $\nabla u$ and $\nabla e_1$ as a consequence of (\ref{phi}).

From all these properties we can show that the principal null directions, $\ell_{\pm}= n \pm \tilde{e}_1$, are null shear-free geodesics, and then these metrics belong to the class considered in \cite{Stewart-Ellis} (see also \cite{Kramer, Wainwright}). The spatially-homogeneous cosmologies in this family will be considered in subsection \ref{subsec-TypeD-nonconstant}.


\section{Spatially homogeneous cosmologies admitting an invariant frame}
\label{sec-regular-SHC}

Now, we can make use of the results in the above two sections to undertake the ideal labelling of the spatially-homogeneous cosmologies admitting a Riemann-frame. 
For the sake of simplicity, we extend the terminology used for the type D case as follows: the perfect fluid solutions admitting a Riemann-frame will be named {\em regular cosmologies}. In the previous two sections, we have presented how to obtain the connection tensor $H$ associated with a Riemann-frame of these solutions. To facilitate our presentation, the following definition will be useful.
\begin{definition} \label{def-main}
We refer as {\em main connection tensor} of a regular cosmology to the tensor $H$ given in:
\begin{itemize}
\item[(i)]
Proposition {\em \ref{propo-typeI}} for solutions of type I.
\item[(ii)]
Proposition {\em\ref{propo-N-III-II}} for solutions of types N, III or II.
\item[(iii)] Propositions {\em\ref{propo-caso1}}, {\em \ref{propo-constant}} or {\em \ref{propo-nonconstant}} for regular type D solutions.
\end{itemize}
\end{definition}

In the case of regular cosmologies we can apply the characterization theorems stated in section \ref{sec-dimension}. We can find two different situations. First, the spacetime is homogeneous (with a transitive G$_4$ group of isometries) and it admits a subgroup G$_3$ on space-like three-dimensional orbits. Second, the spacetimes admits a maximal simply-transitive group G$_3$ of isometries on space-like orbits. These two cases will be considered in the following subsections.


\subsection{Homogeneous spatially-homogeneous cosmologies with a G$_4$}
\label{subsec-regular-SHC-G4}

In this case, we can apply proposition \ref{propo-G4} and obtain the characterization of the regular cosmologies admitting a transitive group G$_4$.  
To end the study of this case, we must analyze if the spacetime admits a subgroup G$_3$ on space-like three-dimensional orbits, and determine the Bianchi type of this subgroup.

The homogeneous perfect fluid solutions can be interpreted as a dust solution with cosmological constant since the energy density and the pressure are constants. These solutions with positive matter density (that is $q>0$) are all known (see \cite{Kramer} and references therein). The ones admitting a maximal transitive G$_4$ split into two different families distinguished by the unimodular or non-unimodular nature of the G$_4$ group \cite{MacCallum-G4}. 

Farnsworth and Kerr \cite{Farnsworth-Kerr} showed that the unimodular solutions correspond with the Ozsv\'ath  \cite{Ozsvath_b} classes I, II and III. As stated in \cite{Kramer}, metrics in class I are spatially-homogeneous and admit a G$_3$ of Bianchi type IX on S$_3$. They can be labelled by the invariant condition $r  > 0$. If $r <0$, the solutions of class II and III obtained by Ozsv\'ath  \cite{Ozsvath_b} are not spatially homogeneous (see \ref{apen-Ozsvath} for more details).  

The non-unimodular family was also found by Ozsv\'ath \cite{Ozsvath_b} and they fall into three different classes, depending on the sign of the Ricci invariant:
\begin{equation}\label{invaB}
\hat{\beta} = 1 - \frac{(3q-5r)(3q-r)(q+r)}{4 (q-r)^3} ,
\end{equation}
where $r$ and $q$ are given in (\ref{fluper-definitions}). As stated in \cite{Kramer}, when $\hat{\beta} \geq 0$ the spacetime is spatially-homogeneous and admits a Bianchi type VI$_h$ on S$_3$. However, if $\hat{\beta}<0$ the G$_3$ subgroups act necessarily on time-like orbits.

In order to obtain a full IDEAL labelling of these geometries we must write the unimodular condition in terms of Riemann concomitants. Again, we can use the main connection tensor $H$. Indeed, from (\ref{nabla-killing}) we can obtain the commutator $[\xi_a, \xi_b]$ of two Killing vectors, and consequently, the group structure constants $C_{ab}^c$. Then, if $\{\xi_a\}$ is a frame of Killing vectors, we obtain $C_{ab}^c = H_{ba}^{\ c} -H_{ab}^{\ c}$. Thus, $2 A_a \equiv  C_{ab}^b = H_{ba}^{\ b} = (\tr H)_a$ and, since the unimodular case can be characterized by $A_a = 0$ \cite{Kramer}, we can state:
\begin{proposition} 
Let $H$ be the connection tensor of a spacetime admitting a simply-transitive G$_4$. The group is unimodular if, and only if, $\tr H =0$.
\end{proposition}
From all the results above we get:
\begin{theorem} \label{theo-regularG4}
A regular cosmology is homogeneous with a G$_4$ if, and only if, its main connection tensor $H$ (definition {\em \ref{def-main}}) fulfills 
condition G$_4$ given in {\em (\ref{G4})}. Moreover, it is a dust solution with cosmological constant and positive matter density if, and only if, $q>0$. Furthermore:
\begin{itemize}
\item[(i)]
If $\tr H =0$ (unimodular), the spacetime is spatially-homogeneous if, and only if, $r >0$. Then, a subgroup G$_3$ of Bianchi IX on S$_3$ exists.
\item[(ii)]
If $\tr H \not=0$ (non-unimodular), the spacetime is spatially-homogeneous if, and only if, $\hat{\beta} \geq 0$, where $\hat{\beta}$ is given in {\em (\ref{invaB})}. Then, a subgroup G$_3$ of Bianchi type $VI_h$ exists. 
\end{itemize}
\end{theorem}
%


\subsection{Spatially-homogeneous cosmologies with a maximal G$_3$}
\label{subsec-regular-SHC-G3}

Now, we can apply proposition \ref{propo-G3} and obtain the characterization of the regular cosmologies admitting a maximal group of isometries G$_3$. Moreover, proposition \ref{propo-G3causal} gives us the condition for space-like orbits. Then, we can state:
\begin{theorem} \label{theo-regularG3}
A regular cosmology is spatially-homogeneous with a G$_3$ as maximal group if, and only if, its main connection tensor $H$ (definition {\em \ref{def-main}}) fulfills condition G$_3$ given in {\em (\ref{G3})}, and
\begin{equation} \label{m}
m^2 < 0 , \qquad  m_\lambda = C^{[1]}_{\lambda \rho \mu \nu}v^\rho V^{\mu \nu} ,
\end{equation}
where $v$ is a vector and $V$ a two-form such that $m \not=0$.
\end{theorem}
It is worth remarking that (\ref{m}) is a generic condition, valid for any regular cosmology. However, for the type D cases studied in subsection \ref{subsec-nonconstant}, this causal condition, $m^2 < 0$, necessarily holds as a consequence of (\ref{mu}). 

On the other hand, we can also determine the Bianchi type of the cosmological model by using the main connection tensor. Note that the normal vector and the induced metric on the three-dimensional orbits are given, respectively, by
\begin{equation} \label{n-zeta}
n = |m^2|^{-1/2} m , \qquad \tilde{\gamma} = g + n \otimes n .
\end{equation}
We can apply to the metric $\tilde{\gamma}$ our results in \cite{FS-G3} on homogeneous three-dimensional Riemannian spaces. In that paper we determine the Bianchi type in terms of the structure tensor $Z$ associated to the isometry group. This $Z$ is characterized by the condition $\stackrel{3}{\nabla}\! \xi=\ \stackrel{3}{*}
i(\xi) Z$ for every Killing vector $\xi$, where $\stackrel{3}{\nabla}$ and $\stackrel{3}{*}$ denote the covariant derivative and the Hodge dual operator of the Riemannian metric $\tilde{\gamma}$ \cite{FS-G3}. From here, and taking (\ref{nabla-killing}) into account, we can obtain $Z$ in terms of the main connection tensor $H$  as:
\begin{equation} \label{Z}
Z_{\alpha \beta} = - \frac{1}{2}\tilde{\gamma}^{\lambda}_{\alpha} \,
{H_{\lambda}}^{\mu \nu} \eta_{\mu \nu \beta \rho} \, n^{\rho} .
\end{equation}
Then, we can state:
\begin{theorem} \label{theo-Bianchi-type}
For the spatially-homogeneous cosmologies of theorem {\em \ref{theo-regularG3}}, the structure tensor of the three-dimensional orbits is given by {\em (\ref{Z})}. Then, the Bianchi type can be obtained with the algorithm presented in subsection {\em 2.2} of reference {\em \cite{FS-G3}}. 
\end{theorem}
%


\section{Singular spatially-homogeneous cosmologies}
\label{sec-singular-SHC}

If the spacetime does not admit a Riemann-frame, the results in the previous section do not apply. In these cases, we shall see that a multiply transitive group of isometries exists and a further analysis will be necessary to know whether the solution is spatially-homogeneous. This means to study if the group (or a subgroup) acts transitively on orbits S$_3$. We analyze separately the conformally flat case and the singular type D solutions.


\subsection{Type O}
\label{subsec-TypeO}

All the conformally flat perfect fluid solutions are known and classified into two different families attending to the vanishing or not of the fluid expansion. When $\theta =0$, they are generalizations of the Schwarzschild interior, and when $\theta \neq0$ they are the Stephani universes \cite{Stephani_a} (see also \cite{Kramer}).

In the second case, the metric is spatially-homogeneous if, and only if, the fluid velocity $u$ is geodesic, or equivalently, the fluid is barotropic. This means that the metric is the FLRW solution, which admits a G$_6$ on three-dimensional space-like orbits.

When $u$ is expansion-free, the metric is given by \cite{Kramer}:
\begin{eqnarray}
\dif s^2 = - F^2 \dif t^2 + \frac{1}{1- C^2 r^2} \dif r^2 + r^2 (\dif
\theta^2 + \sin^2 \theta \dif \phi^2) ,\\
\hspace{-22mm} F \equiv r f_1(t) \sin\theta \, \sin
\phi + r f_2(t) \sin\theta \, \cos\phi+ r f_3(t) \cos\theta+ f_4(t)
\sqrt{1\!-\!C^2 r^2} - C^{-1} ,
\end{eqnarray} 
where $C$ is a non vanishing constant and $f_i(t)$ are arbitrary real functions. Moreover, $2 C F= \rho +p =q$.

If $F$ is a constant, then the spacetime is the Einstein static universe, which admits a maximal group G$_7$ with a subgroup G$_6$ on three-dimensional space-like orbits \cite{Kramer}.

If $F$ is not a constant, then $\dif F \not=0$ is orthogonal to the orbits when the solution is spatially-homogeneous. Moreover, $F_t = u(F)$ is an invariant and then, necessarily, $\dif F_t  \wedge \dif F=0$. This condition leads to $F_t=0$ after a straightforward reasoning. Then, the metric is static and the group has three-dimensional time-like orbits. 

We summarize the results on type O metrics in the following:
\begin{theorem} 
\label{theo-typeO}
The spatially-homogeneous conformally flat perfect fluid metrics are  characterized by the condition $\dif r \wedge \dif q =0$. When $\dif(3q+r)\not=0$ they are the FLRW universes, which admit a G$_6$ on three-dimensional orbits S$_3$. When $\dif(3q+r)=0$ the solution is the static Einstein universe, which is homogeneous admitting a G$_7$, with a G$_6$ on S$_3$. 
\end{theorem}
%


\subsection{Singular Type D with constant algebraic scalars}
\label{subsec-TypeD-constant}

Now equations (\ref{nablau-nablae}) hold, and then the Ricci identities for the vector $u$ lead to $s w =0$. The case $s=w =0$ makes the Weyl tensor to become zero, and then its not compatible with  type D. Then, two different situations arise, namely, $s=0$ and $s \not=0$.

When $s =0$, we obtain the Gödel solution \cite{Godel}, which admits a transitive G$_5$ containing a G$_3$ of type III on space-like
three-dimensional orbits \cite{Kramer}.

When $s \not=0$, we obtain the metric of the Kompaneets-Chernov-Kantowski-Sachs (KCKS) family \cite{KS, Kom-Cher} (see \ref{KS}), with $k=0$ and $\theta =0$, that admits a transitive G$_5$ with a G$_3$ of type I on space-like three-dimensional orbits. This way, we have the following:
\begin{theorem}
\label{theo-typeD-constant}
The singular type D perfect fluid solutions with constant algebraic invariants are spatially-homogeneous, and they admit a transitive G$_5$. Moreover:
\begin{itemize}
\item[(i)] If $s=0$ they are the Gödel solution, which admits a G$_3$ of Bianchi type III on S$_3$. 
\item[(ii)]
If $s \not=0$ they are the KCKS metrics with $k=0=\theta$, which admit a G$_3$ of Bianchi type I on S$_3$.
\end{itemize}
\end{theorem}
%


\subsection{Singular Type D with a non-constant algebraic scalar}
\label{subsec-TypeD-nonconstant}

Now, a necessary condition for the existence of a group G$_3$ is (\ref{lambda}). Moreover, equations (\ref{nabla-e1}) hold, and then the Ricci identities for the vector $\tilde{e}_1$ lead to $s_1 w_1 =0$. Then, we have three possible cases that have been analyzed in \ref{apen-typeD-nonconstant}, and the results can be summarized in the theorem below. In it, we use these scalar invariants $s_1$ and $w_1$, and a new one $\zeta$, which take the following expressions in terms of Riemann concomitants already defined:
\begin{equation} \label{s1-w1}
\begin{array}{ll}
4 s_1^2 \equiv \tilde{s}_1^2, & \qquad  (\tilde{s}_1)^{\alpha} = \tilde{\gamma}^{\lambda \mu} \nabla_{\lambda} (\tilde{P}_1)_{\mu \nu} (\tilde{P}_1)^{\nu \alpha}, \\[2mm] 
4 w_1^2 = \tilde{\omega}_1^2 , & \qquad (\tilde{\omega}_1)^{\alpha} = - \eta^{\alpha \beta \lambda \mu} \nabla_{\beta} (\tilde{P}_1)_{\lambda \nu} (\tilde{P}_1)^{\nu}_{\ \mu} .
\end{array}
\end{equation} 
\begin{equation} \label{zeta}
\ \ \zeta \equiv 4 w_1^2 + \frac{1}{3}\tilde{s} (\tilde{\theta} - \tilde{s}) - 3 \psi_R . 
\end{equation}
\begin{theorem}
\label{theo-typeD-nonconstant}
The singular type D perfect fluid solutions with non-constant algebraic invariants are spatially-homogeneous provided that conditions {\em (\ref{mu})}, {\em (\ref{dmu2})} and {\em (\ref{lambda})} hold. Then, they admit a G$_4$ on S$_3$. Moreover:
\begin{itemize}
\item[(i)] If $s_1 =w_1=0$, they are the KCKS solutions (with $k^2 + \theta^2 \not=0$). When $\zeta = 0$ one has $k=0$ (respectively, $\zeta < 0$ one has $k=-1$) and they admit a G$_3$ on $S_3$ of Bianchi type VII$_0$ (respectively, type III). 
\item[(ii)]
If $w_1=0$ and $s_1 \not=0$, they admit a G$_3$ of Bianchi type V on S$_3$.
\item[(iii)] 
If $w_1\not=0$ and $s_1=0$, they admit a G$_3$ on S$_3$, of Bianchi type: II if $\zeta=0$, VII if $\zeta <0$, or IX if $\zeta>0$.
\end{itemize}
\end{theorem}
%


\section{Summary in algorithmic form}
\label{sec-algoritmes}

Theorems stated in the above two sections offer an IDEAL labelling of the spatially-homogeneous cosmologies. These statements involve scalar and tensorial quantities depending only on the curvature tensor. Consequently, an algorithm can be performed that enables us to distinguish different classes and to test if a specific metric in a class defines a spatially-homogeneous cosmology. 

For the sake of understanding and accuracy, we present the algorithm in several flow diagrams that we explain in the following subsections. The sole input data of these diagrams are the Ricci and Weyl tensors, which can be obtained from the metric tensor in an arbitrary coordinate system.


\begin{figure}[h]

\hspace*{0.5cm} \setlength{\unitlength}{0.9cm} {\small \noindent
\begin{picture}(0,13)

\thicklines

\put(5,12){\line(1,0){1.5}} \put(5,12 ){\line(-1,1){0.5}}
\put(5,13){\line(-1,-1){0.5}} \put(6.5,13 ){\line(-1,0){1.5}}
\put(6.5,13 ){\line(0,-1){1}} \put(4.9,12.35){$\Ric, \  {\cal W}$}

\put(2,13.1){\line(-2,-1){1.25}} \put(2,13.1){\line(2,-1){1.25}}
\put(2,11.85){\line(2,1){1.25}} \put(2,11.85){\line(-2,1){1.25}}
\put(1.28,12.34){Theo.$\,$\ref{theo-fluper-energy}}


\put(0.4,11.1){\line(-2,-1){1.25}} \put(0.4,11.1){\line(2,-1){1.25}}
\put(0.4,9.85){\line(2,1){1.25}} \put(0.4,9.85){\line(-2,1){1.25}}
\put(-0.25,10.35){${\cal W} = 0$}

\put(3.4,11.1){\line(-2,-1){1.25}} \put(3.4,11.1){\line(2,-1){1.25}}
\put(3.4,9.85){\line(2,1){1.25}} \put(3.4,9.85){\line(-2,1){1.25}}
\put(2.74,10.35){${\cal W}^2 = 0$}


\put(6.4,11.1){\line(-2,-1){1.25}} \put(6.4,11.1){\line(2,-1){1.25}}
\put(6.4,9.85){\line(2,1){1.25}} \put(6.4,9.85){\line(-2,1){1.25}}
\put(5.74,10.35){${\cal W}^3 = 0$}
\put(9.4,11.1){\line(-2,-1){1.25}} \put(9.4,11.1){\line(2,-1){1.25}}
\put(9.4,9.85){\line(2,1){1.25}} \put(9.4,9.85){\line(-2,1){1.25}}
\put(8.6,10.32){$6 b^2 \neq a^3$}

\put(13,11.35){\line(-2,-1){1.75}} \put(13,11.35){\line(2,-1){1.75}}
\put(13,9.6){\line(2,1){1.75}} \put(13,9.6){\line(-2,1){1.75}}
\put(11.6,10.33){$a {\cal W}^2\!\!\! -\!\frac{a^2}{3} \!{\cal
G}\!\neq \!b {\cal W}   $}


\put(0.4,8.8){{\oval(0.9,0.6)}} \put(0.25,8.65){O}


\put(3.4,8.8){{\oval(0.9,0.6)}} \put(3.24,8.65){N}

\put(2.6,7.5){\framebox{$H\!=\!H_N$}}

  \put(6.4,8.8){{\oval(0.9,0.6) }}\put(6.18,8.65){III}

\put(5.6,7.5){\framebox{$H\!=\!H_{III}$}}

 \put(9.4,8.8){{\oval(0.9,0.6) }}\put(9.32,8.65){I}

\put(8.6,7.5){\framebox{$H\!=\!H_I$}}

 \put(13,8.8){{\oval(0.9,0.6) }}\put(12.85,8.65){II}

 \put(12.1,7.5){\framebox{$H\!=\!H_{II}$}}

\put(16,8.8){{\oval(0.9,0.6) }}\put(15.85,8.65){D}

\put(15.35,6.5){Fig.$\!$ \ref{figure-3}}
\put(-0.1,6.5){Fig.$\!$ \ref{figure-2}}

\put(3.4,7.3){\line(0,-1){0.5}} \put(2.9,6.5){Fig.$\!$ \ref{figure-4}}

\put(6.4,7.3){\line(0,-1){0.5}} \put(5.9,6.5){Fig.$\!$  \ref{figure-4}}

\put(9.4,7.3){\line(0,-1){0.5}} \put(8.9,6.5){Fig.$\!$  \ref{figure-4}}

\put(13,7.3){\line(0,-1){0.5}} \put(12.5,6.5){Fig.$\!$  \ref{figure-4}}

\put(16,10.5){\vector(0,-1){1.4}} \put(16,10.5){\line(-1,0){1.25}}
\put(15.4,10.6){no}

\put(0.4,12.46){\vector(0,-1){1.36}}

 \put(0.75,12.47){\line(-1,0){0.36}}

\put(2,11.84){\vector(0,-1){0.3}}

\put(1.8,11.14){\#}

 \put(4.5,12.49){\vector(-1,0){1.3}}
\put(0.4,9.86){\vector(0,-1){0.76}}
\

\put(3.4,9.86){\vector(0,-1){0.76}}
\put(6.4,9.86){\vector(0,-1){0.76}}\put(9.4,9.86){\vector(0,-1){0.76}}
\put(13,9.6){\vector(0,-1){0.5}}

\put(13,8.5){\line(0,-1){0.58}}

\put(16,8.5){\line(0,-1){1.58}}

\put(0.4,8.5){\line(0,-1){1.58}} \put(3.4,8.5){\line(0,-1){0.58}}

\put(6.4,8.5){\line(0,-1){0.58}} \put(9.4,8.5){\line(0,-1){0.58}}

\put(1.66,10.47){\vector(1,0){0.5}}

\put(4.66,10.47){\vector(1,0){0.5}}

\put(7.66,10.47){\vector(1,0){0.5}}

\put(10.66,10.47){\vector(1,0){0.6}}

\put(-0.4,11.7){yes}  \put(2.2,11.65){no}

\put(-0.4,9.45){yes}

\put(2.6,9.45){yes} \put(5.6,9.45){yes} \put(8.6,9.45){yes}
\put(12.3,9.35){yes}


\put(1.7,10.6){no}\put(4.7,10.6){no} \put(7.7,10.6){no}
\put(10.7,10.6){no}
\end{picture} }
\vspace*{-5.5cm} \caption{This flow diagram tests whether the metric is a perfect fluid solution and determines the main connection tensor for Petrov-Bel types N, III, II, and I.} \label{figure-1}
\end{figure}
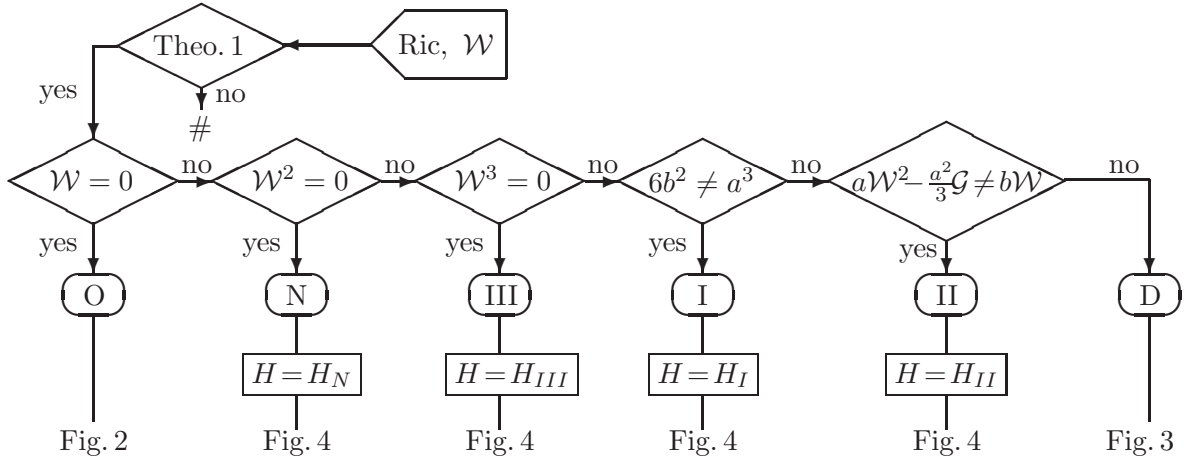


\subsection{Perfect fluid condition and connection tensor of Petrov-Bel types N, III, I and II}
\label{subsec-Type-PB-flow}

An initial requirement that must be imposed on the metric tensor is that it defines a perfect fluid solution. Thus, a first step in the characterization algorithm is to impose the conditions in theorem \ref{theo-fluper-energy} by using the Ricci concomitants $r$, $q$ and $Q$. 

Figure \ref{figure-1} shows a flow diagram with this first step. We use the number sign $\#$ to indicate that the metric does not fulfill the perfect fluid conditions. We can also find in this figure an algorithm that enables us to distinguish the Petrov-Bel type of the metric. It is similar (with minor changes) to the one presented in \cite{FMS-Weyl}. The notation on the self-dual Weyl tensor is given after lemma \ref{lemma-PE}, and the Weyl scalar invariants $a$ and $b$ are defined in proposition \ref{propo-typeI}.

Figure \ref{figure-1} also shows that for Petrov-Bel types N, III, II and I the corresponding main connection tensor $H_N$, $H_{III}$, $H_{II}$ and $H_I$ can be obtained (propositions \ref{propo-N-III-II} and \ref{propo-typeI}). Once obtained, we can use the flow diagram in figure \ref{figure-4} to determine the isometry group. The analysis of types O and D are presented, respectively, in the flow diagrams of figures \ref{figure-2} and \ref{figure-3}.


\subsection{Type O spatially-homogeneous cosmologies}
\label{subsec-regular-flow}

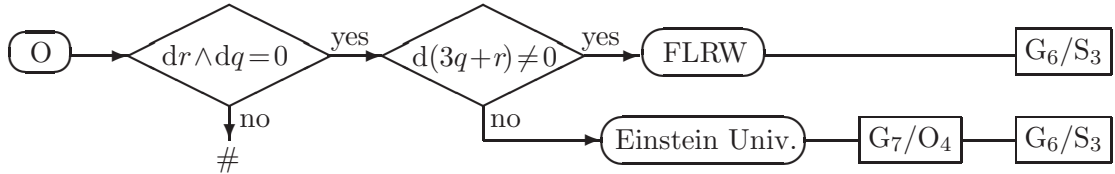
\begin{figure}[h]
\vspace*{7mm}
 \setlength{\unitlength}{0.9cm} {\small \noindent
 \hspace*{5mm}
\begin{picture}(0,13)

\thicklines


\put(0.2,12.9){{\oval(0.9,0.6)}} \put(0.05,12.75){O}

\put(0.65,12.85){\vector(1,0){0.85}}

\put(3,13.6){\line(-2,-1){1.5}} \put(3,13.6){\line(2,-1){1.5}}
\put(3,12.1){\line(2,1){1.5}} \put(3,12.1){\line(-2,1){1.5}}
\put(2,12.72){$\dif r\! \wedge\! \dif q\! =\! 0$}
\put(4.45,12.85){\vector(1,0){0.8}}

\put(6.75,13.6){\line(-2,-1){1.5}} \put(6.75,13.6){\line(2,-1){1.5}}
\put(6.75,12.1){\line(2,1){1.5}} \put(6.75,12.1){\line(-2,1){1.5}}
\put(5.7,12.7){$\dif (3q\!+\!r\!) \!\neq\! 0$}
\put(8.25,12.85){\vector(1,0){0.8}}


\put(10,11.6){\oval(3,0.7)} \put(8.7,11.46){Einstein Univ.}

\put(10,12.9){\oval(1.8,0.7)}\put(9.4,12.75){FLRW}

\put(12.3,11.5){\framebox{G$_7$/O$_4$}}
\put(14.61,11.5){\framebox{G$_6$/S$_3$}}
\put(14.61,12.8){\framebox{G$_6$/S$_3$}}

\put(6.75,12.1 ){\line(0,-1){0.5}}
\put(6.75,11.6){\vector(1,0){1.74}}
 \put(3,12.1){\vector(0,-1){0.5}}  \put(2.8, 11.2){\#}

\put(10.9,12.85){\line(1,0){3.72}}

\put(11.5,11.6){\line(1,0){0.8}} \put(13.8,11.6){\line(1,0){0.8}}

\put(3.15,11.8){no}

\put(6.85,11.8){no}

\put(8.2,13){yes}

\put(4.5,13){yes}
\end{picture} }
\vspace*{-10cm} \caption{This flow diagram characterizes the spatially-homogeneous cosmologies of Petrov-Bel type O.} \label{figure-2}
\end{figure}

The spatially-homogeneous cosmologies of Petrov-Bel type O are analyzed in subsection \ref{subsec-TypeO}. The figure \ref{figure-2} summarizes theorem \ref{theo-typeO} by means of a flow diagram. The sole input data are the two Ricci scalar invariants $r$ and $q$. Now, the number sign $\#$ indicates that the metric is not spatially-homogeneous.

In this figure \ref{figure-2}, and in the following ones \ref{figure-3} and \ref{figure-4}, the output information shows (inside a box and in the form G$_n$/S$_3$) the dimension $n$ of the maximal group acting on three-dimensional space-like orbits. When the space is homogeneous, the dimension $m$ of the maximal group is also shown (inside another box and in the form G$_m$/O$_4$).


\subsection{Type D: singular cosmologies and connection tensor of regular classes}
\label{subsec-Type-D-flow}

Figure \ref{figure-3} shows a flow diagram that analyzes the Petrov-Bel type D case. Again, we use the number sign $\#$ to indicate that the metric is not spatially-homogeneous.

On the one hand, the diagram shows the different classes of regular cosmologies that follow from our study. At every step we need to define new Riemann concomitants that allow us to obtain the main connection tensor $H$ of a specific class. The expression of these invariants can be found in section \ref{sec-typeD}, and the role that they play in the flow diagram is explained in propositions \ref{propo-constant} and \ref{propo-nonconstant}. The case considered in proposition \ref{propo-caso1} leads to $H_1$, the three cases considered in proposition \ref{propo-constant} lead to $H_2$, $H_\sigma$ and $H_3$, and the three cases considered in proposition \ref{propo-nonconstant} lead to $H_4$, $H_{\tilde{\sigma}}$ and $H_5$. Once $H$ is obtained, we can use the flow diagram in figure \ref{figure-4} to determine the isometry group.


\begin{figure}[h]
\vspace{13mm}
\hspace*{0.25cm} \setlength{\unitlength}{0.9cm} {\small \noindent
\begin{picture}(0,15.5)

\thicklines

\put(3,16.15){{\oval(1.2,0.8) }}\put(2.86,16){{\bf{D}}}
\put(3.6,16.12){\vector(1,0){1.2}}

\put(6,16.75){\line(-2,-1){1.25}} \put(6,16.75){\line(2,-1){1.25}}
\put(6,15.5){\line(2,1){1.25}} \put(6,15.5){\line(-2,1){1.25}}
\put(5.4,16 ){$\delta_1 =0$}


\put(6,15){\line(-3,-1){2.15}} \put(6,15){\line(3,-1){2.15}}
\put(6,13.55){\line(3,1){2.15}} \put(6,13.55){\line(-3,1){2.15}}
\put(4.52,14.14){$\dif r \!= \dif q\! =\dif \psi\! =\!0$}



\put(6,13){\line(-2,-1){1.25}} \put(6,13){\line(2,-1){1.25}}
\put(6,11.75){\line(2,1){1.25}} \put(6,11.75){\line(-2,1){1.25}}
\put(5.3,12.25 ){$E_2 =0$}

\put(1.5,12.3){\framebox{$H\!=\!H_2$}} \put(-0.4,12.3){Fig.$\!$
\ref{figure-4}}
\put(6,11.2){\line(-2,-1){1.25}} \put(6,11.2){\line(2,-1){1.25}}
\put(6,9.95){\line(2,1){1.25}} \put(6,9.95){\line(-2,1){1.25}}
\put(5.4,10.45 ){$\delta_\sigma =0$}

\put(1.5,10.5){\framebox{$H\!=\!H_\sigma$}} \put(-0.4,10.5){Fig.$\!$
\ref{figure-4}}

\put(6,9.4){\line(-2,-1){1.25}} \put(6,9.4){\line(2,-1){1.25}}
\put(6,8.15){\line(2,1){1.25}} \put(6,8.15){\line(-2,1){1.25}}
\put(5.32,8.65 ){$E_3=0$}

\put(1.5,8.7){\framebox{$H\!=\!H_3$}} \put(-0.4,8.7){Fig.$\!$
\ref{figure-4}}

\put(5.2,7.2){\framebox{G$_5$/O$_4$}}

\put(6,6.65){\line(-2,-1){1.25}} \put(6,6.65){\line(2,-1){1.25}}
\put(6,5.4 ){\line(2,1){1.25}} \put(6,5.4){\line(-2,1){1.25}}
\put(5.5,5.88 ){$s =0$}

\put(2.5,5.85 ){\framebox{Gödel}}

\put(-0.4,5.85){\framebox{B III/S$_3$}}

\put(1.7,4.7 ){\framebox{\,KCKS, \,$k=\!0\!=\theta\,$\phantom{$a^b$\hspace*{-3.7mm}}}}

\put(-0.4,4.7){\framebox{B I/S$_3$}}


\put(10,14.85){\line(-2,-1){1.25}} \put(10,14.85){\line(2,-1){1.25}}
\put(10,13.6){\line(2,1){1.25}} \put(10,13.6){\line(-2,1){1.25}}
\put(9.25,14.1 ){Eq. $\!$(27)}



\put(10,13){\line(-2,-1){1.25}} \put(10,13){\line(2,-1){1.25}}
\put(10,11.75){\line(2,1){1.25}} \put(10,11.75){\line(-2,1){1.25}}
\put(9.45,12.25 ){$\delta_4 =0$}

\put(12.8 ,12.3){\framebox{$H=H_4$}} \put(15.5,12.3){Fig.$\!$
\ref{figure-4}}

\put(12.8 ,16){\framebox{$H=H_1$}} \put(15.5,16){Fig.$\!$ \ref{figure-4}}

\put(16.34,14.1){\#}

\put(16.34,10.5){\#}

\put(12.8 ,8.7){\framebox{$H=H_{\tilde{\sigma}}$}}
\put(15.5,8.7){Fig.$\!$ \ref{figure-4}}

\put(12.8 ,6.86){\framebox{$H=H_5$}} \put(15.5,6.86){Fig.$\!$
\ref{figure-4}}



\put(10,11.2){\line(-2,-1){1.25}} \put(10,11.2){\line(2,-1){1.25}}
\put(10,9.95){\line(2,1){1.25}} \put(10,9.95){\line(-2,1){1.25}}
\put(9.3,10.45 ){Eq.$\!$ (31)}




\put(10,9.4){\line(-2,-1){1.25}} \put(10,9.4){\line(2,-1){1.25}}
\put(10,8.15){\line(2,1){1.25}} \put(10,8.15){\line(-2,1){1.25}}
\put(9.4,8.6 ){$\tilde{\delta}_{\sigma} =0$}


\put(10,7.6){\line(-2,-1){1.25}} \put(10,7.6){\line(2,-1){1.25}}
\put(10,6.35){\line(2,1){1.25}} \put(10,6.35){\line(-2,1){1.25}}
\put(9.4,6.85 ){$E_5 = 0$}

\put(10,5.75){\line(-2,-1){1.25}} \put(10,5.75){\line(2,-1){1.25}}
\put(10,4.5){\line(2,1){1.25}} \put(10,4.5){\line(-2,1){1.25}}
\put(9.3,5 ){Eq.$\!$ (36)}

\put(11.2,5.13){\vector(1,0){4.8}} \put(16.34, 5){\#}

\put(9.3,3.5){\framebox{G$_4$/S$_3$}}

\put(10,2.75){\line(-2,-1){1.25}} \put(10,2.75){\line(2,-1){1.25}}
\put(10,1.5){\line(2,1){1.25}} \put(10,1.5){\line(-2,1){1.25}}
\put(9.4,2){$w_1= 0$}

\put(12.2,2.13){\line(1,0){2.8}} \put(11.2,2.13){\vector(1,0){1}}

\put(12.2,0.95){\line(1,0){2.8}}

\put(12.2,3.35){\line(1,0){2.5}}

\put(12.2,0.95){\line(0,1){2.4}}

\put(15,2.05){\framebox{\ B II/S$_3$}}

\put(15,0.85){\framebox{B IX/S$_3$}}

 \put(14.7, 3.25){\framebox{B VIII/S$_3$}}

\put(13.2,1.15){$\zeta >0$}

\put(13.2,2.35){$\zeta =0$}

\put(13.2,3.55){$\zeta <0$}

\put(-0.4,2.65){\framebox{B III/S$_3$}}

\put(-0.4,1.45){\framebox{B VII$_0$/S$_3$}}

\put(3.2,2.15){\line(1,0){0.45}}

\put(5.77,2.13){\vector(-1,0){0.65}}

\put(8.77,2.13){\vector(-1,0){0.55}}

\put(3.65,2.05){\framebox{KCKS}}

\put(3.2,1.45){\line(0,1){1.4}}

\put(3.2,2.85){\line(-1,0){1.76}}

\put(3.2,1.45){\line(-1,0){1.42}}

\put(2,3.01){$\zeta <0$}

\put(2,1.65){$\zeta=0$}

\put(7,2.75){\line(-2,-1){1.25}} \put(7,2.75){\line(2,-1){1.25}}
\put(7,1.5){\line(2,1){1.25}} \put(7,1.5){\line(-2,1){1.25}}
\put(6.45,2.0){$s_1= 0$}

\put(7,1.5){\vector(0,-1){0.6}}

\put(6.2,0.45){\framebox{B V/S$_3$}}

\put(6,15.5){\vector(0,-1){0.5}}

\put(6,13.55){\vector(0,-1){0.55}}

\put(6,11.78){\vector(0,-1){0.6}}

\put(6,9.95 ){\vector(0,-1){0.55}}

\put(6,8.16 ){\vector(0,-1){0.5}}

\put(6,6.98 ){\line(0,-1){0.33}}

\put(6,5.4 ){\line(0,-1){0.6}}


\put(6,4.8){\vector(-1,0){0.65}} \put(1.7,4.8){\line(-1,0){0.6}}

\put(4.8,6.02){\vector(-1,0){0.94}} \put(2.5,6){\line(-1,0){1.06}}

\put(7.2,16.12){\vector(1,0){5.6}}

\put(14.55,16.12){\line(1,0){0.6}}

\put(8.1,14.27){\vector(1,0){0.68}}
\put(11.23,14.24){\vector(1,0){4.8}}

\put(11.2,12.38){\vector(1,0){1.6}}
\put(14.55,12.38){\line(1,0){0.6}}

\put(4.75,12.38){\vector(-1,0){1.6}}
\put(1.5,12.38){\line(-1,0){0.6}}

\put(4.75,10.58){\vector(-1,0){1.6}}
\put(1.5,10.58){\line(-1,0){0.6}}

\put(11.23,10.58){\vector(1,0){4.8}}

\put(4.75,8.78){\vector(-1,0){1.6}} \put(1.5,8.78){\line(-1,0){0.6}}

\put(11.2,8.78){\vector(1,0){1.6}} \put(14.6,8.78){\line(1,0){0.6}}

\put(11.2,6.98){\vector(1,0){1.6}} \put(14.6,6.98){\line(1,0){0.6}}

%

\put(10,13.62){\vector(0,-1){0.6}}

\put(10,11.78){\vector(0,-1){0.6}}

\put(10,9.95 ){\vector(0,-1){0.55}}

\put(10,8.16 ){\vector(0,-1){0.57}}

\put(10,6.38 ){\vector(0,-1){0.63}}

\put(10,4.5 ){\vector(0,-1){0.55}}
 \put(10,3.27){\line(0,-1){0.51}}


\put(8.1,16.3){no} \put(8.1,14.5){no}

\put(11.5,14.4){no} \put(11.5,12.6){no}

\put(11.5,10.8){no} \put(11.5,9){no}\put(11.5,7.2){no}
\put(11.3,2.35){no} \put(8.3,2.35){yes}

\put(5.3,2.35){yes} \put(7.2,1.1){no}

\put(4.2,12.6){no} \put(4.2,10.8){no} \put(4.2,9){no}
\put(11.5,7.2){no}

\put(4.2,6.2){no}

\put(6.2,15.2){yes}

\put(6.2,13.2){yes} \put(10.2,13.2){yes}

\put(6.2,11.4){yes} \put(10.2,11.4){yes}

\put(6.2,9.6){yes} \put(10.2,9.6){yes}

\put(6.2,7.85){yes} \put(6.2,5){yes}

\put(10.2,7.8){yes}

\put(10.2,6){yes}

\put(11.5,5.3){no}

\put(10.2,4.2){yes}

\end{picture} }

\vspace*{-2mm}
\caption{This flow diagram determines the main connection tensor for the regular type D cosmologies. Moreover, for singular type D cosmologies, it determines the dimension of the isometry group and distinguishes the Bianchi type of the cases when a G$_3$ acts on three-dimensional space-like orbits.}
\label{figure-3}
\end{figure}


On the other hand, the flow diagram also summarizes theorems \ref{theo-typeD-constant} and \ref{theo-typeD-nonconstant}, which study the spatially-homogeneous singular type D cosmologies. To discriminate the different cases, we need to define new invariant scalars in each step: the $s$ defined in (\ref{nablau-nablae}) and $s_1$, $w_1$ and $\zeta$ defined in (\ref{s1-w1}) and (\ref{zeta}). All the metrics (except in the case of a KCKS metric with $\zeta>0$) admit a subgroup G$_3$ on space-like three-dimensional orbits. The Bianchi type of these subgroups appears inside a box as output data of the algorithm.


\subsection{Regular spatially-homogeneous cosmologies}
\label{subsec-regular-flow}

\begin{figure}[h]
\vspace{3mm} \hspace*{0.5cm} \setlength{\unitlength}{0.9cm} {\small
\noindent
\begin{picture}(0,15.3)
\thicklines \put(0.9,15){H} \put(1,14.65){\vector(0,-1){0.54}}
\put(1.6,14.95){\line(-2,-1){0.6}} \put(0.4,14.95){\line(2,-1){0.6}}
\put(1.6,14.95){\line(0,1){0.6}} \put(0.4,14.95){\line(0,1){0.6}}
\put(0.4,15.55){\line(1,0){1.2}}
\put(1,14.1){\line(-2,-1){1.25}} \put(1,14.1){\line(2,-1){1.25}}
\put(1,12.85){\line(2,1){1.25}} \put(1,12.85){\line(-2,1){1.25}}
\put(0.35,13.35){Eq. $\!$(5)} \put(1,12.85){\vector(0,-1){1.5}}
\put(3.65,13.4){{\framebox{G$_4$/O$_4$}}}
\put(7,14.1){\line(-2,-1){1.25}} \put(7,14.1){\line(2,-1){1.25}}
\put(7,12.85){\line(2,1){1.25}} \put(7,12.85){\line(-2,1){1.25}}
\put(6.2,13.35){$\,\tr H\!=\!0$} \put(7,12.85){\vector(0,-1){0.35}}

\put(10.75,14.24){\line(-2,-1){1.5}}
\put(10.75,14.24){\line(2,-1){1.5}}
\put(10.75,12.74){\line(2,1){1.5}}
\put(10.75,12.74){\line(-2,1){1.5}} \put(9.7,13.4){$r\!>\!0, \, q
\!>\! 0$}


\put(5.55,11.76){\vector(-1,0){0.75}} \put(4.23, 11.65){\#}

\put(8.2,13.48){\vector(1,0){1.07}}

\put(12.2,13.48){\vector(1,0){2.2}}

\put(10.75,14.88){\vector(1,0){2}}

\put(10.75,14.25){\line(0,1){0.64}}

\put(12.9, 14.8){\#}

\put(8.5,11.75){\vector(1,0){5.8}}

\put(7,12.5){\line(-2,-1){1.5}} \put(7,12.5){\line(2,-1){1.5}}
\put(7,11){\line(2,1){1.5}} \put(7,11){\line(-2,1){1.5}}
\put(6.1,11.63){$\tilde{\beta}\! \geq \!0, q \!>\!0$}
%
\put(14.3,11.66){\framebox{B VI$_h$/S$_3$}}

\put(12.3,13.7){yes} \put(10.9,14.36){no}

\put(8.7,12 ){yes}

\put(5.17,12){no}

\put(14.4,13.4){{\framebox{B IX\,/S$_3$}}}
\put(1,11.35){\line(-2,-1){1.75}} \put(1,11.35){\line(2,-1){1.75}}
\put(1,9.6){\line(2,1){1.75}} \put(1,9.6){\line(-2,1){1.75}}
\put(-0.12,10.35){Eqs.$\!$ (6)$\!$ (39)}
\put(1,9.6){\vector(0,-1){0.4}} \put(0.83,8.75){\#}
\put(3.65,10.4){{\framebox{G$_3$/S$_3$}}}

\put(5.06,10.5){\vector(1,0){4.04}}

\put(2.75,10.48){\vector(1,0){0.9}}
\put(7.25,10.25){\line(-4,-1){1.2}}
\put(7.25,10.25){\line(4,-1){1.2}} \put(6.04,9.45){\line(0,1){0.5}}
 \put(8.45,9.45){\line(0,1){0.5}}
 \put(6.04,9.45){\line(1,0){2.4}}
\put(6.37,9.64){$Z\!\equiv\!Z(H)$}
 \put(7.25,10.25){\line(0,1){0.24}}
\put(9,10.35){\Big[{Algorithm Ref.$\,$\cite{FS-G3}$\,$}\Big]}
\put(13.7,10.35){\framebox{Bianchi type
\phantom{$a^b$\hspace*{-4.5mm}}}}

%
%
\put(5.16,13.48){\vector(1,0){0.64}}
\put(2.19,13.48){\vector(1,0){1.48}}
\put(12.92,10.5){\line(1,0){0.75}}
\put(2.3,13.7){yes}
\put(8.3,13.7){yes}
\put(1.15,12.1){no} \put(1.15,9.3){no} \put(2.8,10.75){yes}
\end{picture} }
\vspace*{-8.0cm} \caption{This flow diagram characterizes the
regular spatially-homogeneous cosmologies once the main connection tensor $H$ is known.} \label{figure-4}
\end{figure}
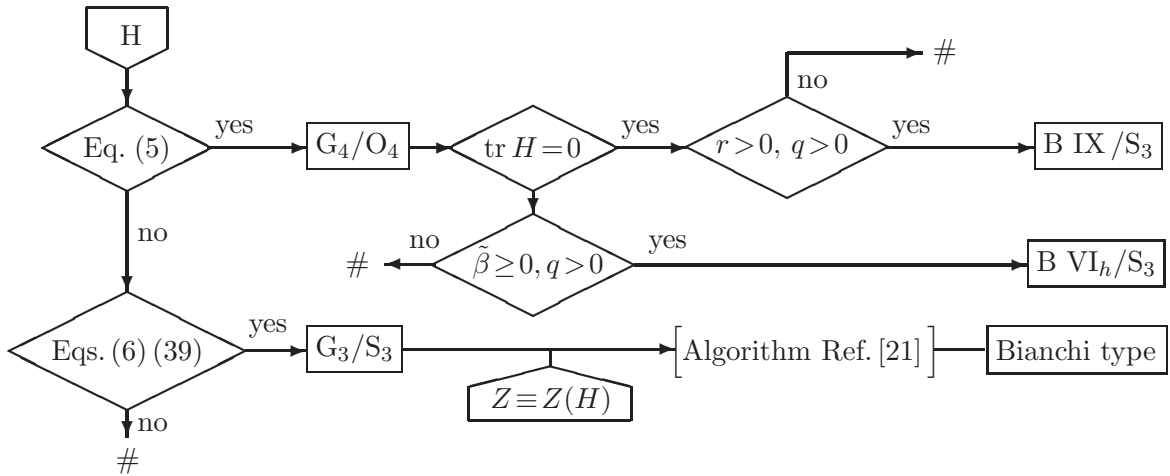

The spatially-homogeneous cosmologies admitting a Riemann frame have been characterized in section \ref{sec-regular-SHC}, and more specifically in theorems \ref{theo-regularG4} and \ref{theo-regularG3}. Figure \ref{figure-4} summarizes these results in a flow diagram where the main connection tensor $H$ is the only initial input data. This $H$ is the output data in figure \ref{figure-1} (Petrov-Bel types N, III, II and I) and in figure \ref{figure-2} (regular type D cosmologies). 

When the space-time is homogeneous (G$_4$/O$_4$) we only consider the dust solutions with cosmological constant and positive matter density (see theorem \ref{theo-regularG4}). Then, we use $H$ and the Ricci invariants $r$ and $q$ given in (\ref{fluper-definitions}) and $\tilde{\beta}$ given in (\ref{invaB}) to know if a G$_3$ on $S_3$ exists and what is its Bianchi type. Again, the number sign $\#$ indicates that the metric is not spatially-homogeneous.

When the maximal group is a G$_3$, we must define the tensor $Z = Z(H)$ given in (\ref{Z}) to determine the corresponding Bianchi type by using an algorithm given in \cite{FS-G3}. 

\section{Implementing the algorithm on {\em xAct}. Two examples}
\label{sec-exemples}

The algorithmic nature of the IDEAL characterizations makes them particularly suitable to be implemented in a formal calculation computer program. One such formal calculation program is \textit{xAct}, a suite of \textit{Mathematica} packages for tensor computer algebra \cite{xAct}. It has attracted a considerable amount of users, as can be seen from the extensive bibliography of articles and theses that have used it (see the articles section in \cite{xAct}). One of the authors of this paper (SM) is developing, in collaboration with A. Garc\'{\i}a-Parrado, an \textit{xAct} package called \textit{xIdeal} that implements IDEAL characterizations and other algorithms devised by our group. It will be published elsewhere when finished, but it can be found in its current unfinished version at \cite{xIdeal}.

In this section, we will use \textit{xIdeal} to apply the algorithm presented in this paper to particular solutions as examples.


\subsection{Bianchi type I cosmologies}
\label{subsec-Bianchi-I}

Let us consider now the general expression of the metrics admitting a three-dimensional group of isometries of Bianchi type I \cite{ellis-maar-mac}:
\begin{equation}
\dif s^2 \, = \, - \dif t^2 + \ell_1^2(t) \, \dif x^2 + \ell_2^2(t) \, \dif y^2 + \ell_3^2(t) \, \dif z^2 \, .
\end{equation}
This example can serve as a test for our algorithm. Conditions in (\ref{fluper-conditions-A2}) give us the field equations for $\ell_i(t)$ that should be solved in order to obtain perfect fluid solutions, which are always non-tilted ($u=-\dif t$). Regardless of that, using the corresponding function of \textit{xIdeal} we get that these metrics are, generically, of Petrov Type I. This allows the programm to obtain its connection tensor $H_I$ by applying proposition \ref{propo-typeI} (see figure \ref{figure-1}) and use it to implement the algorithm in figure \ref{figure-4}. With that, we get that Eq. (\ref{G4}) is not fulfilled but Eqs. (\ref{G3}) and (\ref{m}) are. Therefore, we check that, indeed, the dimension of the isometry group is three. At this point, we will also use this example to test the algorithm proposed in \cite{FS-G3} to determine the Bianchi type of the group. To do so, we use \textit{xIdeal} to obtain the structure tensor $Z$ given in theorem \ref{theo-Bianchi-type} and we get that it vanishes. Therefore \cite{FS-G3}, the $G_3$ is indeed of Bianchi type I .


\subsection{The homogeneous Ozsv\'ath solutions of classes II and III}
\label{subsec-Farnsworth-Kerr}

Let us consider the Ozsv\'ath solutions of classes II and III as 
given in \cite{Kramer}, 
\begin{equation} \label{FK-Kramer}
\dif s^2 \, = \, \hat{a}^2 [ (1 - \hat{s}) (\omega^1)^2 + (1 + \hat{s}) (\omega^2)^2 + ( \dif u + \lambda \omega^3)^2 - 2(\omega^3)^2 ] \, , 
\end{equation}
with $\lambda =0$ for class III, and $\lambda^2= 1- 2 \hat{s}^2$ for class II. The parameters $\hat{a}$ and $\hat{s}$ are constants, and $\{\omega^1, \omega^2, \omega^3\}$ are the reciprocal group of a Bianchi type VIII, namely \cite{Kramer}, $\omega^1 = \dif x - \sinh{y} \, \dif z$, $\omega^2 = \cos{x} \, \dif y - \sin{x} \cosh{y} \, \dif z$, and $\omega^3 = \sin{x} \, \dif y + \cos{x} \cosh{y} \, \dif z$.  

We can compute the Ricci invariants $q$ and $Q$ given in (\ref{fluper-definitions}) and we get that: (i) For $\lambda = 0$, $q \, Q(\omega^3, \omega^3) = 0$. Thus, the second condition in (\ref{fluper-conditions-A2}) does not hold and, consequently, metric (\ref{FK-Kramer}) is not a perfect fluid solution. In fact, the Ricci tensor has a triple eigenvalue and the first condition in (\ref{fluper-conditions-A2}) holds, but the eigenvector associated with the simple eigenvalue is space-like. (ii) For $\lambda \neq 0$, the first condition in (\ref{fluper-conditions-A2}) does not hold, that  is, the Ricci tensor does not have a triple eigenvalue and, consequently, metric (\ref{FK-Kramer}) is not a perfect fluid solution.

Thus, there is a mistake in the transcription that is made in \cite{Kramer} of the results by Ozsv\'ath. Indeed, in the notation of \cite{Kramer}, the solutions of classes II and III take actually the expression \cite{Ozsvath_b, Farnsworth-Kerr}: 
\begin{equation} \label{Ozsvath-II-III}
\dif s^2 \, = \, \hat{a}^2 [- 2(\omega^1)^2 + (1 + \hat{s}) (\omega^2)^2 + ( \dif u + \lambda \omega^1)^2 + (1 - \hat{s})(\omega^3)^2 ] \, , 
\end{equation}
with $\lambda$, $\hat{a}$, $\hat{s}$ and $\{\omega^1, \omega^2, \omega^3\}$ as given after (\ref{FK-Kramer}). Now, we obtain that both conditions in (\ref{fluper-conditions-A2}) are fulfilled, and thus, these metrics are perfect fluid solutions. Moreover, as stated in \cite{Ozsvath_b, Farnsworth-Kerr}, they define dust solutions with cosmological constant and positive matter density if: $\hat{s}^2 <1$ when $\lambda =0$ (class III) and $1/4< \hat{s}^2 <1/2$ when $\lambda \not=0$ (class II).

On the other hand, using the corresponding function of \textit{xIdeal} (see figure \ref{figure-1}) we get that metrics (\ref{Ozsvath-II-III}) are of Petrov Type I when $\lambda \not=0$, and of type D when $\lambda =0$.

For class III Ozsv\'ath metrics, we use the {\em xIdeal} function that implements the algorithm to obtain the connection tensor $H$ of a type D solution given in figure \ref{figure-3}. Then, we get that (i) the fluid unit velocity $u$ is in the principal plane of the Weyl tensor, $\delta_1 = 0$; (ii) all the algebraic scalar invariants are constant, that is, condition (\ref{drqpsi}) holds; (iii) the rotation vector $\omega$ is collinear to $e_1$, $\omega \wedge e_1 = 0$, so that, $E_2=0$; (iv) the shear tensor $\sigma$ is algebraically general, $\delta_\sigma \neq 0$. Therefore, this is the case (ii) in proposition \ref{propo-constant}, and $H= H_{\sigma}$ can be computed as in lemma \ref{lemma-PE}. 

For class II Ozsv\'ath metrics, we could obtain the main connection tensor $H$ applying proposition \ref{propo-typeI}. Alternatively, we can get $H$ by using the function that implements expression (\ref{defH}) in terms of the Weyl principal frame $\{e_{0}, e_{1}, e_{+}, e_{-}\}$. This frame can be determined as indicated in \cite{FMS-Weyl} by using the corresponding function of {\em xIdeal} and we obtain:
\begin{equation}
\hspace{-10mm} e_0 = \hat{a}\sqrt{2}\, \omega_1 , \quad e_1 = \hat{a}\, (\dif u + \lambda \omega_1), \quad e_{\pm} =  \frac{\hat{a}}{\sqrt{2}}\, [\sqrt{1+ \hat{s}}\, \omega_2 \pm \sqrt{1- \hat{s}}\,  \omega_3]. 
\end{equation}

In the two cases, once we get the connection tensor $H$, we apply the algorithm proposed in figure \ref{figure-4}. From it, we get that the dimension of the isometry group is four and that $\tr H =0$, that is, the metric admits an unimodular $G_4$. We also have $q>0$, $r<0$. All these invariant conditions characterize the Ozsv\'ath solutions of classes II and III. Nevertheless, they are not spatially-homogeneous solutions, contrary to what Ozsv\'ath \cite{Ozsvath_b} and \cite{Kramer} claim (see \ref{apen-Ozsvath}).

It is worth remarking that solutions (\ref{Ozsvath-II-III}) were obtained under the hypothesis of a positive matter density. The determination of the solution without this condition is still an open problem that will lead to new solutions. For example, the metric given by:
\begin{equation} \label{FK}
\dif s^2 \, = \, \hat{a}^2 [ 2 (\omega^1)^2 + (1 + \hat{s}) (\omega^2)^2 + 2 \dif u^2 - (1 - \hat{s})(\omega^3)^2 ] \, , 
\end{equation}
is a dust solution with cosmological constant, which has similar properties as the class III Ozsv\'ath metric, but with a negative matter density.


\section{Discussion and work in progress}
\label{sec-discussion}

In this paper we have obtained an IDEAL characterization of the spatially-homogeneous cosmologies. Our study requires considering an invariant classification of the cosmologies defined by intrinsic and explicit conditions on the Ricci and Weyl tensors. For instance, the Petrov-Bel type or the possible alignment of the fluid velocity with the principal plane of a type D Weyl tensor play an important role. For each class, we have obtained the necessary and sufficient conditions for a metric to be a spatially-homogeneous perfect fluid solution. Our invariant study enables us to perform algorithms which can be efficiently implemented in formal calculation computer programs.

We have requested that the metrics be a perfect fluid solution, that is, the Ricci tensor admits a space-like three-dimensional eigenspace, without any complementary physical constraints. Only in subsection \ref{subsec-regular-SHC-G4} we have considered a positive matter density. Nevertheless, the basic results and the structure of our approach remain valid when we impose the energy conditions (see characterization theorem in \cite{fs-SSST-Ricci, CFS-CC}) or when we require that the solution could be interpreted as a perfect fluid in local thermal equilibrium (see characterization theorem in \cite{fs-SSST-Ricci}).

Our results enable us to know if a metric tensor, given in an arbitrary coordinate system, defines a spatially-homogeneous cosmology. Nevertheless, we can quote other subsidiary applications. Thus, in looking for new solutions of the field equations, it could be helpful to restrict the search to one of the (invariant) classes considered in our paper. It might also be interesting to apply our algorithm to the families of solutions already known to determine which of our classes they belong to. This study will offer an IDEAL labelling of these solutions and enables us to better understand their intrinsic geometric properties. The possible relation of the asymptotic dynamic behavior of the models (see \cite{Horwood-2003, Hervik-2010} and references therein) with our classification is a query to be studied.

On the other hand, there are some other open questions that could be analyzed taking into account our results. The compatibility of the tilted models with the Bianchi type and with the kinematics of the fluid flow has been analyzed in \cite{King-Ellis}. Here we have seen that only one of the type D models studied in \ref{apen-typeD-nonconstant} (Bianchi type V) is tilted. The study of the compatibility of each of our classes with the tilted condition is a work in progress.

Another question to be analyzed is the study of isometries admitted by the induced metric $\tilde{\gamma}$. As a consequence of the results in \cite{FS-G3}, the analysis of the Ricci tensor of $\tilde{\gamma}$ enables us to determine if the orbits admit a G$_4$ or a $G_6$ as maximal group. For instance, in Bianchi type I models the orbits are flat three-dimensional spaces and $\tilde{\gamma}$ admits, necessarily, a G$_6$.

\ack 
This work has been supported by the Generalitat Valenciana Project CIAICO/2022/252, the Spanish Ministerio de Ciencia, Innovaci\'on, Project PID2019-109753GB-C21/AEI/10.13039/501100011033, and the Plan Recuperaci\'on, Transformaci\'on y Resiliencia, project ASFAE/2022/001, with funding from European Union NextGenerationEU (PRTR-C17.I1). S.M. acknowledges financial support from the Generalitat Valenciana (grant CIACIF/2021/028). 


\appendix


\section{Singular Type D spatially-homogeneous cosmologies with a non-constant algebraic invariant}
\label{apen-typeD-nonconstant}

Here, we consider the spatially-homogeneous metrics that appear when a non-constant algebraic invariant exists, that is, condition (\ref{mu}) holds, and when all the Riemann concomitants that enable us to obtain a main connection tensor vanish (see subsection \ref{subsec-TypeD-nonconstant}). In this case, the pair $\{n, \tilde{e}_1\}$ generates the principal plane and fulfills (\ref{nabla-e1}), with the gradient of all the involved scalars collinear with $n$. Then, for an orthonormal frame $\{e_2, e_3\}$ in the space-like principal plane, we obtain that a 1-form $h$ exists such that
\begin{equation}
\begin{array}{l}
 \nabla e_2 = \frac{1}{3} (\tilde{\theta} - \tilde{s}) e_2 \otimes
n - (s_1 e_2 + w_1 e_3 ) \otimes \tilde{e}_1 + h \otimes e_3, 
\\[2mm]
\nabla e_3 = \frac{1}{3} (\tilde{\theta} - \tilde{s}) e_3 \otimes n
+ (w_1 e_2 - s_1 e_3) \otimes \tilde{e}_1 - h \otimes e_2 .
\end{array}
\end{equation}
Note that we have the freedom of a rotation of angle $\alpha$ in the plane $\{e_2, e_3\}$, and then, $h$ changes to $h + \dif \alpha$, that is, $h$ is defined up to a gradient. On the other hand, the Ricci identities for $\tilde{e}_1$ imply $w_1 s_1 =0$. Then several cases can be outlined.


\subsection{$w_1=0=s_1$} 
\label{app-0-0}

In this case, the metric belongs to the Kompaneets-Chernov-Kantowski-Sachs family of solutions \cite{KS, Kom-Cher}:
\begin{equation} \label{KS}
\hspace{-20mm} \dif s^2 = - \dif t^2 + X^2(t) \dif z^2 + Y^2(t) \dif \Sigma^2, \quad \ \dif \Sigma^2 = \frac{1}{[1+\frac{k}{4}(x^2 + y^2)]^2} (\dif x^2 + \dif y^2) ,
\end{equation}
where $k=0,\pm 1$, and the fluid flow is aligned, $u = n =- \dif t$. This family covers all the KCKS metrics (\ref{KS}) except the case $\tilde{\theta}=0= k$, which has constant algebraic invariants and has been considered in subsection \ref{subsec-TypeD-constant}. The invariant $k$ is determined by the sign of the scalar $\zeta$ defined in (\ref{zeta}), $k = sign[\zeta]$.  
All these solutions are spatially homogeneous admitting a G$_4$ on S$_3$, However, when $k=1$ it does not admit a G$_3$ on three-dimensional orbits. On the other hand, if $k=0$, a G$_3$ of Bianchi type VII$_0$ acts on S$_3$ and when $k=-1$, we have a G$_3$ of Bianchi type III \cite{Kramer}.


\subsection{$w_1=0\neq s_1$} 
Now, the Ricci identities state that the Weyl eigenvalue is real ($\Psi_I =0$) and $\dif h=0$, which means that a rotation in the plane $\{e_2, e_3\}$ can be performed leading to $h=0$. The exterior system for the tetrad $\{n,  \tilde{e}_1, e_2 , e_3 \}$ implies that the metric takes the expression:
\begin{equation} \label{metric-Ap}
g = - \dif t^2 + X^2(t) \omega_1 + Y^2(t) (\omega_2
\otimes \omega_2 + \omega_3 \otimes \omega_3),
\end{equation}
where $n = - \dif t$, and $\{ \omega_1, \omega_2 , \omega_3 \}$ satisfy the exterior system of the reciprocal group of a Bianchi type V. Then, coordinates $\{x, y, z\}$ exist such that
$\omega_1 = \dif x $, $\omega_2 = e^x \dif y $, $\omega_3 = e^x \dif z$ \cite{Kramer}. Moreover, the field $\xi_4= \partial_x - y \partial_y - z \partial_z$ is also a Killing vector and, consequently, the metric admits a G$_4$ on S$_3$ with a G$_3$ on S$_3$ of Bianchi type V. In this case we have a tilted model, $u \not= n$.


\subsection{$w_1 \neq 0 = s_1$}  
In this case, the fluid flow is aligned, $u=n$, and $\{n, \tilde{e}_1 , e_2, e_3 \}$ satisfy
\begin{eqnarray}
\hspace{-20mm} \dif n = 0, \quad \dif \tilde{e}_1 = \frac{1}{3}(2 \tilde{s}\!
+\! \tilde{\theta}) \tilde{e}_1 \wedge n + 2 w_1 e_3 \wedge e_2, \quad \dif (h + w_1 \tilde{e}_1) = \zeta\ e_3 \wedge e_2,  \label{A4}  \\
\hspace{-23mm} \dif e_2 = \frac{1}{3}(\tilde{\theta}\! -\! \tilde{s}) e_2 \wedge n +
(h\!+\! w_1 \tilde{e}_1) \wedge e_3 , \quad \dif e_3 =
\frac{1}{3}(\tilde{\theta}\! -\! \tilde{s}) e_3 \wedge n - (h\!+\! w_1
\tilde{e}_1) \wedge e_2 , 
\end{eqnarray}
where $\zeta$ is given in (\ref{zeta}). From these equations we obtain $\dif \zeta =0$. Then,
it is adequate to outline two subcases.


\subsubsection{$\zeta =0$.} 
Now, we can perform a rotation in the plane $\{e_2, e_3\}$ to get $h+w_1 e_1 =0$. Then, the metric takes the expression (\ref{metric-Ap}), where $n = - \dif t$, and $\{ \omega_1, \omega_2 , \omega_3 \}$ satisfy the exterior system of the reciprocal group of a Bianchi type II. Then, coordinates $\{x, y, z\}$ exist such that $\omega_1 = \dif x - z \dif y$, $\omega_2 =  \dif y $, $\omega_3 = e^x \dif z$ \cite{Kramer}. Moreover, the field $\xi_4= \frac{1}{2} (z^2 - y^2) \partial_x + z \partial_y - y \partial_z$ is also a Killing vector and, consequently, the metric admits a G$_4$ on S$_3$ with a G$_3$ on S$_3$ of Bianchi type II.


\subsubsection{$\zeta \neq 0$.} 
In this case, the metric tensor takes the expression (\ref{metric-Ap}), where $n = - \dif t$, and $\{ \omega_1, \omega_2 , \omega_3 \}$ satisfy the exterior system of the reciprocal group of a Bianchi type VIII when $\zeta <0$ (respectively, type IX when $\zeta>0$). Then, coordinates $\{x, y, z\}$ exist such that $\omega_1= \dif x - \sinh{y} \dif z$, $\omega_2=\cos{x} \dif y - \sin{x} \cosh{y} \dif z $, $\omega_3 = \sin{x} \dif y + \cos{x} \cosh{y} \dif z$ if $\zeta <0$ (respectively, $\omega_1= \dif x + \sin{y} \dif z$, $\omega_2=\cos{x} \dif y - \sin{x} \cos{y} \dif z $, $\omega_3 = \sin{x} \dif y + \cos{x} \cos{y} \dif z$ if $\zeta >0$) \cite{Kramer}. Moreover, in both cases, the field $\xi_4=\partial_x$ is also a Killing vector and, consequently, the metric admits a G$_4$ on S$_3$, with a G$_3$ on S$_3$ of Bianchi type VII if $\zeta <0$, or Bianchi type IX if $\zeta >0$. 


\section{The homogeneous Ozsv\'ath solutions of class II and III are not spatially homogeneous}
\label{apen-Ozsvath}

The Ozsv\'ath solutions of class II and III given in (\ref{Ozsvath-II-III}) admit a simply transitive G$_4$ with reciprocal group $\{\omega^1, \omega^2, \omega^3, \omega^4\}$, where $\omega^4 = \dif u$. Then, four independent Killing vectors $\{\xi_1,\xi_2,\xi_3,\xi_4\}$ exist such that:
\begin{equation} \label{xi_}
\hspace{-10mm}[\xi_3, \xi_2] = \xi_1, \quad [\xi_3, \xi_1] = \xi_2, \quad [\xi_1, \xi_2] = \xi_3, \quad  [\xi_i, \xi_4] = 0, \ \ i=1,2,3.
\end{equation}
Thus, G$_4 =$ G$_3$ $\oplus$ G$_1$, G$_3$ being a subgroup of Bianchi type VIII, and it is known \cite{Patera} that this G$_4$ also admits a subgroup G$_3$ of Bianchi type III. This means that three Killing vectors $\{X_1,X_2,X_3\}$ exist such that:
\begin{equation} \label{X_}
[X_2, X_1] = [X_2, X_3] = 0 \qquad [X_1, X_3] = X_1.
\end{equation}
Then, from (\ref{xi_}) and (\ref{X_}) we obtain: 
\begin{equation} \label{X_xi_}
\hspace{-10mm} X_1 = \xi_1 + \cos \varphi \, \xi_2 + \sin \varphi \, \xi_3, \quad  X_2 = \xi_4, \quad  X_3 = \sin \varphi \, \xi_2 - \cos \varphi \, \xi_3,
\end{equation}
where $\varphi \in [0, 2 \pi[$ is a real parameter. 

The causal character of the orbits of the Bianchi III depend on the sign of $Y^2$, where $Y = *(X_1 \wedge X_2 \wedge X_3)$. A long but straightforward calculation leads to $Y^2 >0$ and, consequently, the subgroup G$_3$ of Bianchi type III has time-like orbits. 

This result contradicts that by Ozsv\'ath \cite{Ozsvath_b} (also quoted in \cite{Kramer}) on the existence of a group of Bianchi type III with space-like orbits in the solution of classes II and III.


\end{document}